\documentclass[twocolumn,english,aps,prb]{revtex4-1}
\pdfoutput=1
\usepackage[T1]{fontenc}
\usepackage[latin9]{inputenc}
\usepackage[a4paper]{geometry}
\usepackage{amsmath}
\usepackage{amssymb}
\usepackage{graphicx,color}
\usepackage{esint}

\makeatletter

\providecommand{\tabularnewline}{\\}

\usepackage{babel}

\usepackage{babel}

\usepackage{babel}

\usepackage{babel}

\usepackage{babel}

\makeatother

\usepackage{babel}
\begin{document}

\title{A hydrodynamic model approach to the formation of plasmonic wakes in graphene}

\author{A.J. Chaves$^{1}$, N.M.R. Peres$^{1}$, G. Smirnov$^2$, and N. Asger Mortensen$^{3,4,5}$}

\address{$^{1}$Department and Center of Physics, and QuantaLab, University
of Minho, Campus of Gualtar, PT-4710-374, Braga, Portugal}

\address{$^{2}$Department of Mathematics and Applications, and Center of Physics, University
	of Minho, Campus of Gualtar, PT-4710-374, Braga, Portugal}

\address{$^{3}$Center for Nano Optics, University of Southern Denmark, Campusvej
55, DK-5230 Odense M, Denmark}

\address{$^{4}$Center for Nanostructured Graphene, Technical University of
Denmark, Orsteds Plads 343, DK-2800 Kongens Lyngby, Denmark}

\address{$^5$Danish Institute for Advanced Study, University of Southern Denmark, Campusvej 55, DK-5230 Odense M, Denmark}

\begin{abstract}
Using the hydrodynamic
model in the electrostatic approximation, we describe the formation
of graphene surface plasmons when a charge is in motion either perpendicular
or parallel to a graphene sheet. In the first case, the electron-energy
loss (EEL) spectrum of the electron is computed, showing that the
resonances in the spectrum are linked to the frequency of the graphene
surface plasmons. In the second case, we discuss the formation of
plasmonic wakes due to the dragging of the surface plasmons induced
by the motion of the charge. This effect is similar to Coulomb drag between two electron
 gases at a distance from each other. We derive simple expressions for the
electrostatic potential induced by the moving charge on graphene.
We find an analytical expression for the angle of the plasmonic wake valid in two opposite regimes. 
We show that there is a transition
from a Mach-type wake at high speeds to a Kelvin-type wake at low
ones and identify the Froude number for plasmonic wakes.
We show that the Froude number can be controlled 
externally tunning both the Fermi energy in graphene and the
dielectric function of the environment, a situation with no parallel in ship wakes.
 Using EEL we propose a source of graphene plasmons, based on a graphene drum built in a
 metallic waveguide and activated by an electron beam created by the tip of an electronic microscope. 
 We also introduce the notion of a plasmonic billiard.  
\end{abstract}
\maketitle

\section{Introduction}

The hydrodynamic model~\cite{key-27,key-20} for plasmonics is a
macroscopic approach to a microscopic problem, as was well noted by
Cirac\`{i}~\emph{et al.}~\cite{key-14} (see also Ref.~\onlinecite{key-a}).
This model combines Maxwell's equations, Euler's equation of hydrodynamics
supplemented with a term due to the statistical pressure of an electron
gas, and the continuity equation. This set of equations is used for
describing the nonlocal optical response of either a metallic interface~\cite{key-5,key-12,key-21}
or a metallic nano-structure~\cite{key-6,key-d}. The model can be
applied both to 3D~\cite{key-9}, 2D~\cite{key-2,key-26,key-25},
or 1D metallic structures~\cite{key-6,key-10,key-c}. In the latter
case the situation of two metallic nanoparticles in close proximity
is rather important as it allows us to probe electromagnetic interactions
between them down to the separation of few atoms~\cite{key-22,key-17}.
Indeed, whereas an electromagnetic local description of the dimer optical
properties predicts a divergent enhancement of the electromagnetic
energy density in the gap region between the two nanoparticles, a
nonlocal description predicts a reduction of the field enhancement
when the two particle are at atomic distances from each other, in
agreement with the experimental observations~\cite{key-22,key-17}.

In the past six years, graphene has emerged as a new platform for
studying plasmonic effects in the THz and mid-IR, a spectral range
where noble metal plasmons show poor spatial confinement. Since the
hydrodynamic model can be applied to the 2D electron gas~\cite{key-2},
a natural question arises whether graphene, which supports a massless
electron gas, can also be described by the hydrodynamic model. M\"uller~\emph{et
al.} have shown that the massless electron gas in graphene behaves
as a nearly perfect fluid~\cite{key-16} with the electronic motion
described by the Navier\textendash Stokes equation, from which Euler's
equation follows. This model for electronic motion has subsequently
been applied to the characterization of the conductivity of graphene~\cite{key-13}
as well as to the characterization of its plasmonic properties~\cite{key-25}.

One of the merits of the hydrodynamic model~\cite{key-8,key-7} is
permitting the inclusion of nonlocal effects in the plasmonic response
of the very small metallic nano-structures without much computational
burden. Within this model nonlocality appears due to the dependence
of the statistical pressure on the position of the particle, and Coulomb
interactions are included via the coupling of Euler's equation with
Maxwell's equations (or via Poisson's equation in an electrostatic
calculation). Nonlocal effects emerge when the size of the nano-structures
becomes small enough for coarse graining of the electronic charge
no longer holds~\cite{key-22}. These effects have also impact in
the optical properties of metallic gratings~\cite{key-15}. This
happen when the wavelength of the surface plasmon in the metal is
smaller than typical size of the nano-structures. Typically, the condition
$qc/\omega_{p}\gg1$, where $q$, $c$, and $\omega_{p}$ are the
wavenumber of the surface plasmon, the speed of light in vacuum, and
the plasma frequency of the metal, has to be full-filled for nonlocality
to play an important role in the optical spectrum of the system. Physically,
nonlocality arises due to the smearing of the electronic charge when
probed down to the nanoscale~\cite{key-11}. As a consequence, the
screening of the electromagnetic fields become less efficient when
compared to the local calculation prediction. Given this, 
a simple model~\cite{key-21} for a nonlocal metal was introduced. 

\begin{figure}
	\begin{centering}
		\includegraphics[width=8cm]{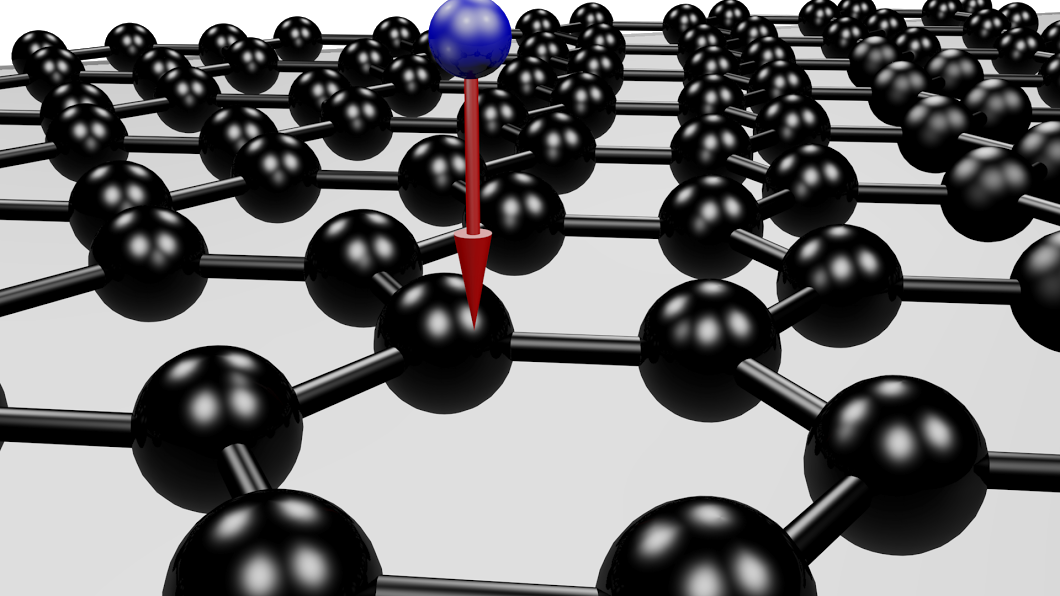} 
		\par\end{centering}
	\begin{centering}
		\includegraphics[width=8cm]{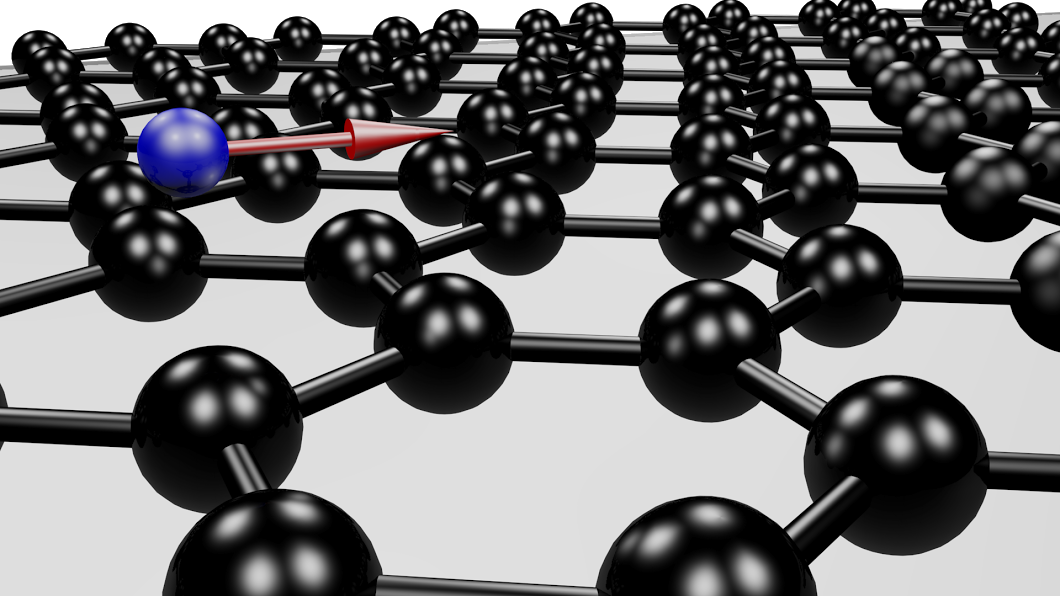} 
		\par\end{centering}
	\caption{System considered in this paper: a charge is moving relatively to graphene either perpendicular (top panel) or parallel (bottom panel) to it. The effect of the interaction of the charge with the electron gas in graphene is studied using the hydrodynamic model, which has built in nonlocal corrections due to the statistical  pressure of the electron gas. The charge induces surface plasmons  in graphene which can be probed by EEL spectroscopy.\label{fig:system}}
\end{figure}

For graphene, the criterium for nonlocality to be important is $qc/(k_{F}v_{F})\gg1$,
with $k_{F}$ and $v_{F}$ being the Fermi wavenumber and Fermi speed
respectively (clearly the system will be highly nonlocal near graphene
neutrality point). For graphene near a metal, the dispersion is strongly
dependent on distance between the two systems. Taking a metal-graphene
distance of about 1.5\,nm, the graphene-metal plasmons can appear
in the mid-IR with a wavenumber of the order of 200\,
${\rm \mu}$m$^{-1}$
(corresponding to a $\lambda_{{\rm spp}}=2\pi/q\approx30$\,nm).
Considering a Fermi energy of graphene of about $E_{F}=0.4$\,eV
the previous condition gives $qc/(k_{F}v_{F})\sim100$, thus placing
graphene in the strong nonlocal regime. In these conditions, this
means that the optical properties of graphene, such as its optical
conductivity, have to be taken as function of both frequency $\omega$
and wave-vector $\mathbf{k}$ in order to account for both temporal
and spatial dispersion. Recently it became possible to retrieve experimentally
the nonlocal optical conductivity of graphene using acoustic graphene
plasmons~\cite{key-23}, introducing a test for the accuracy of many-bod
calculations. In this paper, nonlocality in graphene is taken into
account at the level of the hydrodynamic model, considering a quantum pressure term that is space dependent. Basically, hydrodynamics
is the small-wavenumber correction to the local description, and by
carefully expanding the response function emerging from the random-phase
approximation (RPA) we in general find a two-fluid model to account
for both the interband and intraband effects, which both exhibit spatial
dispersion, but with different nonlocal parameters \cite{key-b}.
In this paper we neglect the interband contribution, since in the
frequency window we are interested these type of transitions are blocked
by Pauli's principle.

The remaining of the paper is organized as follows: in Sec.~ \ref{sec:From-Boltzmann-equation}
we present a short derivation to Euler's equation of hydrodynamics
starting from Boltzmann equation; this sets the stage for the next
sections. In Sec.~\ref{sec:Hydrodynamic-model-for} the hydrodynamic
model for graphene is presented, which will be solved in the 
sections to follow. In Sec.~\ref{sec:2D-hydrodynamic-model} we include external
potentials due to charge densities in the model, which sets the basic
equations for describing the effect of moving charges on graphene
electronic properties. In Sec.~\ref{sec:Induced-electric-field-moving-charge}
the effect of moving charges (see figure \ref{fig:system}) in the induction of plasmonic wakes at
the surface of graphene is studied. In order to discuss plasmonic
effects in graphene nano-structures, we consider in Sec. \ref{sec:nano-rectangle}
the excitation of localized graphene plasmons by an electron in motion,
impinging perpendicularly to a nano-rectangle of graphene located
in a metallic waveguide. Finally, in Sec.~\ref{sec:Conclusions}
we present a short overview of the paper, discuss possible 
extensions of this work, and introduce the concepts of Froude
number for plasmonic wakes and of plasmonic billiards.

\section{Hydrodynamic model for doped graphene in the electrostatic regime\label{sec:Hydrodynamic-model-for}}

The hydrodynamic model of the electromagnetic response of an electron
gas couples Euler's equation to Maxwell's equations \cite{key-8,key-7,key-24,key-19}.
To emphasize the 2D nature of the problem, we will in the following
introduce $\mathbf{r}=(\mathbf{r}_{\parallel},z)$ with $\mathbf{r}_{\parallel}$
being a 2D position vector in the plane of the graphene ($z=0$),
while $z$ is in the direction perpendicular to the graphene layer.
If we introduce the density of particles in the gas per unit area
as $n(\mathbf{r_{\parallel}})$ then the mass density is defined as
$\rho(\mathbf{\mathbf{r_{\parallel}}})=mn(\mathbf{\mathbf{r_{\parallel}}})$,
where $m$ is the mass of the particle and $\mathbf{r_{\parallel}}$
is the 2D position vector. In terms of $n(\mathbf{\mathbf{r_{\parallel}}})$
Euler's equation reads (where we have dropped the average symbol for
simplicity) 
\begin{equation}
mn(\mathbf{\mathbf{r_{\parallel}}})\frac{\partial\mathbf{v}}{\partial t}+mn(\mathbf{r_{\parallel}})(\mathbf{v}\cdot\nabla)\mathbf{v}=\mathbf{g}mn(\mathbf{\mathbf{r_{\parallel}}})-\nabla P
\end{equation}
If scattering is allowed, then a term of the form $mn(\mathbf{r})\mathbf{v}/\tau$
must be included in the left hand side of the previous equation, and
where $\tau$ is a relaxation time taking into account the non-conservation
of momentum. The physics of this added term stems for the electron-phonon
and electron-electron interactions. We now consider that there is
an external electric field applied the electron gas. This changes
Euler's equation to

\begin{equation}
mn(\mathbf{\mathbf{r_{\parallel}}})\frac{\partial\mathbf{v}}{\partial t}+mn(\mathbf{\mathbf{r_{\parallel}}})(\mathbf{v}\cdot\nabla)\mathbf{v}=en(\mathbf{\mathbf{r_{\parallel}}})\nabla\phi(\mathbf{\mathbf{r_{\parallel}}},z=0)-\nabla P
\end{equation}
where $e$ is the elementary charge, $\phi(\mathbf{r})=\phi(\mathbf{r_{\parallel}},z)$
is the electrostatic potential, and the electric field follows from
$\mathbf{E=-\nabla\phi(\mathbf{r})}$.

Since we are dealing with an electron gas, the simplest approximation
for the pressure is the statistical pressure encoded in the kinetic
energy of the electron gas. For graphene the determination of the
pressure (force per unit length in 2D) requires the calculation of
the kinetic energy of the non-interacting gas as

\begin{equation}
K_{g}=4A\int\frac{d\theta kdk}{(2\text{\ensuremath{\pi}})^{2}}v_{F}\hbar k=v_{F}\hbar\frac{2}{3\pi}\pi^{3/2}N_{e}^{3/2}A^{-1/2}
\end{equation}
where $N_{e}$ in the total number of electrons, $A$ is the area
of the system, and we have expressed the Fermi momentum in graphene
by $k_{F}=\sqrt{\pi n_{0}}$ , where $n_{0}$ is the 2D particle density.
It follows that the pressure is given by 
\begin{equation}
P=-\frac{\partial K_{g}}{\partial A}=v_{F}\hbar\frac{1}{3\pi}(\pi n_{0})^{3/2}.
\end{equation}
Note that the previous quantity is a 2D pressure. Next we consider
that the pressure in the inhomogeneous gas has the same functional
form in terms of density as that given by the previous equation. Using
this assumption, we can now compute the gradient of the pressure as

\begin{equation}
\nabla P=v_{F}\hbar\frac{1}{2}\sqrt{\pi n(\mathbf{r_{\parallel}})}\nabla n(\mathbf{r_{\parallel}}).
\end{equation}
%where $\mathbf{r}=(\mathbf{r}_{\parallel},z)$ and $\mathbf{r}_{\parallel}$ is a 2D (in graphene) position vector, as already noted. 
Dividing the pressure by the graphene's Drude mass we obtain Euler's
equation-of-motion for graphene as 
\begin{equation}
\frac{\partial\mathbf{v}}{\partial t}+(\mathbf{v}\cdot\nabla)\mathbf{v}=\frac{ev_{F}}{\hbar k_{F}}\nabla\phi(\mathbf{r_{\parallel}},0)-\frac{v_{F}^{2}}{2n(\mathbf{r_{\parallel}})}\nabla n(\mathbf{r_{\parallel}})
\end{equation}
where the electrostatic potential depends on $\mathbf{r}$ but is
evaluated at the point in graphene given by $\mathbf{r}=(\mathbf{r}_{\parallel},0)$.
The other two equations defining the hydrodynamic model in the electrostatic
limit are Poison's equation 
\begin{equation}
\nabla^{2}\phi=-\frac{e}{\epsilon_{0}}[n_{+}-n(\mathbf{r_{\parallel}})]\delta(z)
\end{equation}
where $n_{+}$ is the ionic charge density neutralizing the electron
gas, and the continuity equation 
\begin{equation}
\frac{\partial n(\mathbf{r_{\parallel}})}{\partial t}+\nabla\cdot[n(\mathbf{r_{\parallel}})\mathbf{v}]=0
\end{equation}
stating charge conservation in the graphene sheet.

We now linearize the hydrodynamic equations, assuming $n(\mathbf{r_{\parallel}})\approx n_{0}+n_{1}(\mathbf{r_{\parallel}})$
(we note in passing that for metals we have $n_{0}\gg n_{1}$; for
graphene however this is not the case when the system is near the
neutrality point. In this paper we will be far from this regime) and
$\phi(\mathbf{r_{\parallel}})\approx\phi_{0}(\mathbf{r_{\parallel}})+\phi_{1}(\mathbf{r_{\parallel}})$,
and noting that $\mathbf{v\,}$ is already a linear order quantity.
This leads to the linear hydrodynamic model \begin{subequations}
\begin{align}
\frac{\partial\mathbf{v}}{\partial t} & =\frac{ev_{F}}{\hbar k_{F}}\nabla\phi_{1}(\mathbf{r_{\parallel}},0)-\frac{v_{F}^{2}}{2n_{0}}\nabla n_{1}(\mathbf{r_{\parallel}})\label{eq:Euler_linearized}\\
\nabla^{2}\phi_{1}(\mathbf{r}) & =\frac{e}{\epsilon_{0}}\delta(z)n_{1}(\mathbf{r_{\parallel}},0)\label{eq:poisso_2d_density}\\
0 & =\frac{\partial n_{1}(\mathbf{r_{\parallel}})}{\partial t}+n_{0}\nabla\cdot\mathbf{v}\label{eq:continuity_linear}
\end{align} \label{eq:hydrodinamic_model}
%ASGER: note that I have changed the order, placing the continuity equation last in the row - when using the subequation-numbering, this fits nicely with our later discussion where we tend to only display the first two equations. In this way "a" equations links to each other and so do "b" equations ... 
\end{subequations} Note that the second term on the right-hand-side
of equation (\ref{eq:Euler_linearized}) is proportional to $1/n_{0}$
and therefore can rightfully be considered a correction to the first
term. Indeed, we can rewrite this equation as

\begin{equation}
\frac{\partial\mathbf{v}}{\partial t}=\frac{ev_{F}^{2}}{E_{F}}\nabla\phi_{1}(\mathbf{r_{\parallel}},0)-\frac{\pi}{2}\frac{\hbar^{2}v_{F}^{4}}{E_{F}^{2}}\nabla n_{1}(\mathbf{r_{\parallel}}), \label{eq:wave_equation_vector}
\end{equation}
which shows that the second term on the right-hand-side of this equation
is of higher order in powers of $1/E_{F}$. Also the presence of $\hbar^{2}$
in the second term signals the presence of a correction of quantum
nature.

\subsection{Spectrum of the surface plasmons}

For solving the previous three equations we introduce the Fourier
transform in the plane (note that here $\mathbf{k}$ is the in-plane
2D wavevector)

\begin{align}
\mathbf{v}(\mathbf{r}_{\parallel},t) & =\int\frac{d\omega d\mathbf{k}}{(2\text{\ensuremath{\pi}})^{3}}\mathbf{v}(\mathbf{k},\omega)e^{i(\mathbf{k}\cdot\mathbf{r}_{\parallel}-\omega t)}
\end{align}
and equivalent definitions for the pairs of transforms $[n_{1}(\mathbf{r}_{\parallel},t);n_{1}(\mathbf{k},\omega)]$
and $[\phi_{1}(\mathbf{r}_{\parallel},z,t);\phi_{1}(\mathbf{k},z,\omega)]$.
Using the Fourier transforms in the hydrodynamic equations we obtain
\begin{subequations} 
\begin{equation}
-i\omega\mathbf{v}(\mathbf{k},\omega)=\frac{ev_{F}}{\hbar k_{F}}i\mathbf{k}\phi_{1}(\mathbf{k},0,\omega)-\frac{v_{F}^{2}}{2n_{0}}i\mathbf{k}n_{1}(\mathbf{k},\omega)\label{eq:Euler_momentum_space-1}
\end{equation}
for Euler's equation, 
\begin{equation}
\left(\frac{\partial^{2}}{\partial z^{2}}-k^{2}\right)\phi_{1}(\mathbf{k},z,\omega)=\frac{e}{\epsilon_{0}}\delta(z)n_{1}(\mathbf{k},\omega)\label{eq:Green_function_potential}
\end{equation}
\end{subequations} for Poisson's equation, and 
\begin{equation}
0=-i\omega n_{1}(\mathbf{k},\omega)+n_{0}i\mathbf{k}\cdot\mathbf{v}(\mathbf{k},\omega)\label{eq:continuity_momentum_space-1}
\end{equation}
for the continuity equation. Note that equation~(\ref{eq:Green_function_potential})
is nothing but the Green's function. For obtaining $\phi_{1}(\mathbf{k},z,\omega)$
we assume that $\phi_{1}(\mathbf{k},z,\omega)=Ae^{-kz}$, for $z>0$
and $\phi_{1}(\mathbf{k},z,\omega)=Be^{kz}$ for $z<0$. The coefficients
$A$ and $B$ are determined from the boundary conditions: $A=B$
and $-k(A+B)=\frac{e}{\epsilon_{0}}n_{1}(\mathbf{k},\omega)$, which
imply that 
\begin{equation}
A=-\frac{e}{2k\epsilon_{0}}n_{1}(\mathbf{k},\omega).
\end{equation}
Using the last result in equation~(\ref{eq:Euler_momentum_space-1})
it follows a relation between $\mathbf{v}(\mathbf{k},\omega)$ and
$n_{1}(\mathbf{k},\omega)$. Using this relation in the continuity
equation~(\ref{eq:continuity_momentum_space-1}) we obtain 
\begin{equation}
\hbar^{2}\omega_{{\rm spp}}^{2}=\left[2\alpha E_{F}\hbar ck+\frac{v_{F}^{2}\hbar^{2}}{2}k^{2}\right]\approx2\alpha E_{F}\hbar ck\label{eq:omega_spp}
\end{equation}
where $\alpha$ is the fine structure constant, with the approximate
result valid for realistic ($k<k_{F}$) wave numbers. We have, therefore,
recovered the well known result for the square-root dispersion of
graphene surface plasmons in the electrostatic limit~\cite{key-BOOK}.
This is consistent with a small-wavenumber expansion of the intraband
part of the RPA result.

\section{2D hydrodynamic model in the presence of external potentials\label{sec:2D-hydrodynamic-model}}

In this section we follow Fetter~\cite{key-2} for the calculation
of response of the electron gas to external potentials. Let us consider
the additional presence of external electrostatic forces acting on
the electron gas. This is accounted for adding extra terms to both
the hydrodynamic equation and Poisson's equation. These are modified
as \begin{subequations} 
\begin{align}
\frac{\partial\mathbf{v}}{\partial t}=\frac{ev_{F}}{\hbar k_{F}}\nabla[\phi_{1}(\mathbf{r_{\parallel}},0) & +\phi_{{\rm ex}}(\mathbf{r_{\parallel}},0)]-\frac{v_{F}^{2}}{2n_{0}}\nabla n_{1}(\mathbf{r_{\parallel}})  \label{eq:external_hydrodynamic}\\
\nabla^{2}[\phi_{1}(\mathbf{r})+\phi_{{\rm ex}}(\mathbf{r})] & =-\frac{e}{\epsilon_{0}}\rho_{{\rm ex}}(\mathbf{r})+\frac{e}{\epsilon_{0}}\delta(z)n_{1}(\mathbf{r_{\parallel}},0)
\end{align}
\end{subequations} where $\phi_{{\rm ex}}(\mathbf{r_{\parallel}},z)$
is the external potential due to the external forces, $\rho_{{\rm ex}}(\mathbf{k},z,\omega)$
is the volume density of external charges, and the continuity equation
is unchanged by the presence of the additional potentials. As before,
we introduce the Fourier transform of the different quantities, leading
to \begin{subequations} 
\begin{equation}
-i\omega\mathbf{v}(\mathbf{k},\omega)=\frac{ev_{F}}{\hbar k_{F}}i\mathbf{k}\phi(\mathbf{k},0,\omega)-\frac{v_{F}^{2}}{2n_{0}}i\mathbf{k}n_{1}(\mathbf{k},\omega)\label{eq:hydrodynamic_momentum_space}
\end{equation}
for Euler's equation 
\begin{align}
\left(\frac{\partial^{2}}{\partial z^{2}}-k^{2}\right)\phi(\mathbf{k},z,\omega) & =-\frac{e}{\epsilon_{0}}\rho_{{\rm ex}}(\mathbf{k},z,\omega)\nonumber \\
 & +\frac{e}{\epsilon_{0}}\delta(z)n_{1}(\mathbf{k},\omega)\label{eq:Poisson_rho_external}
\end{align}
\end{subequations} for Poisson's equation, where $\phi(\mathbf{k},z,\omega)=\phi_{1}(\mathbf{k},z,\omega)+\phi_{{\rm ex}}(\mathbf{k},z,\omega)$
{[}the continuity equation is unchanged: $0=-i\omega n_{1}(\mathbf{k},\omega)+n_{0}i\mathbf{k}\cdot\mathbf{v}(\mathbf{k},\omega)${]}.
For solving equation (\ref{eq:Poisson_rho_external}) we use the Green's
function method. The free space Green's function is defined as 
\begin{equation}
\left(\frac{\partial^{2}}{\partial z^{2}}-k^{2}\right)g(\mathbf{k},z-z^{\prime},\omega)=-\delta(z-z^{\prime})\label{eq:Greens_function}
\end{equation}
from where it follows that 
\begin{equation}
\phi(\mathbf{k},z,\omega)=\int dz^{\prime}g(\mathbf{k},z-z^{\prime},\omega)\rho(\mathbf{k},z^{\prime},\omega)
\end{equation}
where 
\begin{equation}
\rho(\mathbf{k},z,\omega)=-\frac{e}{\epsilon_{0}}\left[\rho_{{\rm ex}}(\mathbf{k},z,\omega)-\frac{e}{\epsilon_{0}}\delta(z)n_{1}(\mathbf{k},\omega)\right].
\end{equation}
The solution of equation (\ref{eq:Greens_function}) is well known
and reads $g(\mathbf{k},z-z^{\prime},\omega)=\exp\left(-k\vert z-z^{\prime}\vert\right)/(2k)$.
%I suggest to turn this equation into an in-line equation ...
%\begin{equation} g(\mathbf{k},z-z^{\prime},\omega)=\frac{1}{2k}e^{-k\vert z-z^{\prime}\vert}. \end{equation}
As a consequence, the potential reads 
\begin{align}
\phi(\mathbf{k},z,\omega) & =\int dz^{\prime}\frac{e}{2k\epsilon_{0}}e^{-k\vert z-z^{\prime}\vert}\rho_{{\rm ex}}(\mathbf{k},z^{\prime},\omega)\nonumber \\
 & -\frac{e}{2k\epsilon_{0}}e^{-k\vert z\vert}n_{1}(\mathbf{k},\omega)
\end{align}
which we write compactly as $\phi(\mathbf{k},z,\omega)=\Phi_{{\rm ex}}(\mathbf{k},z,\omega)-\Phi_{2D,1}(\mathbf{k},z,\omega)$
where 
\begin{equation}
\Phi_{{\rm ex}}(\mathbf{k},z,\omega)=\int dz^{\prime}\frac{e}{2k\epsilon_{0}}e^{-k\vert z-z^{\prime}\vert}\rho_{{\rm ex}}(\mathbf{k},z^{\prime},\omega).
\end{equation}
Using this result in the hydrodynamic equation we obtain 
\begin{align}
\omega\mathbf{v}(\mathbf{k},\omega) & =\frac{e}{m}\mathbf{k}[\Phi_{2D,1}(\mathbf{k},0,\omega)-\Phi_{{\rm ex}}(\mathbf{k},0,\omega)]\nonumber \\
 & +\frac{v_{F}^{2}}{2n_{0}}\mathbf{k}n_{1}(\mathbf{k},\omega).
\end{align}
From the previous equation we obtain the velocity which we plug in
the continuity equation, that can be solved for $n_{1}(\mathbf{k},\omega)$,
and the induced potential is given by 
\begin{multline}
\Phi_{{\rm in}}(\mathbf{k},z,\omega)=-\Phi_{2D,1}(\mathbf{k},z,\omega)\\
=\frac{n_{0}e^{3}}{4\epsilon_{0}^{2}m_{g}}\frac{e^{-k\vert z\vert}}{\omega^{2}-\omega_{{\rm spp}}^{2}}\int dz^{\prime}e^{-k\vert z^{\prime}\vert}\rho_{{\rm ex}}(\mathbf{k},z^{\prime},\omega).
\end{multline}
For computing these quantities in real space an inverse Fourier transform
has to be performed.

\section{Induced electrostatic potential due to a moving charge\label{sec:Induced-electric-field-moving-charge}}

Next we want to consider two applications of the central results obtained
in the previous section. We shall consider the calculation of the
induced electrostatic potential $\Phi_{{\rm in}}(\mathbf{r}_{\parallel},z,\omega)$
and induced electric field in graphene, $\mathbf{E}(\mathbf{r}_{\parallel},z,\omega)=-\nabla\Phi_{{\rm in}}(\mathbf{r}_{\parallel},z,\omega)$,
due to a charge $Ze$ moving at the speed $v$. We consider two cases

\begin{subequations}
\begin{align}
\rho_{{\rm ex}}(\mathbf{r}_{\parallel},z,t) & =Z\delta(x)\delta(y)\delta(z-vt),\label{eq:perpendicular_motion}\\
\rho_{{\rm ex}}(\mathbf{r}_{\parallel},z,t) & =Z\delta(x)\delta(y-vt)\delta(z-z_{0}).\label{eq:parallel_motion}
\end{align}
\end{subequations}
Equations (\ref{eq:perpendicular_motion}) and (\ref{eq:parallel_motion})
represent the motion of the moving charge perpendicular to the graphene
plane (piercing it) and the motion of the moving charge parallel to
the graphene plane at a height $z=z_{0}$, respectively. The Fourier
transform in $\mathbf{r_{\parallel}}$ and $t$ of the charge distributions
gives $\rho_{{\rm ex}}(\mathbf{k},z,\omega)=Z/ve^{i\omega z/v}\equiv Z/ve^{izk_{z}}$
and $\rho_{{\rm ex}}(\mathbf{k},z,\omega)=Z\delta(z-z_{0})2\pi\delta(\omega-k_{y}v)$
for equations (\ref{eq:perpendicular_motion}) and (\ref{eq:parallel_motion}),
respectively. In both cases we see a linear relation between the wavenumber
and frequency: $k_{z}=\omega/v$ and $\omega=k_{y}v$ in the perpendicular
and parallel motion, respectively (note, however, that due to lack
of translation invariance along the $z-$direction, $k_{z}$ is not
a conserved quantity; this implies a non-trivial EEL spectrum). For
ease of our later notation, we now introduce a common prefactor $\Phi_{0}\equiv\frac{Zn_{0}e^{3}}{4\epsilon_{0}^{2}m_{g}}\frac{1}{v^{2}}=\frac{Ze}{\epsilon_{0}}\alpha\frac{v_{F}c}{v^{2}}k_{F}$
that will serve to make many integrals dimensionless. Note that $\Phi_{0}$
has units of electric potential. Since $k_{F}=2\pi/\lambda_{F}$,
where $\lambda_{F}$ is the Fermi wavelength, $\Phi_{0}$ can be interpreted
as the average Coulomb energy between two particles in the electron
gas.

\subsection{Motion perpendicular to the graphene sheet \label{sec:perp_motion}}

\begin{figure}
\begin{centering}
\includegraphics[width=8cm]{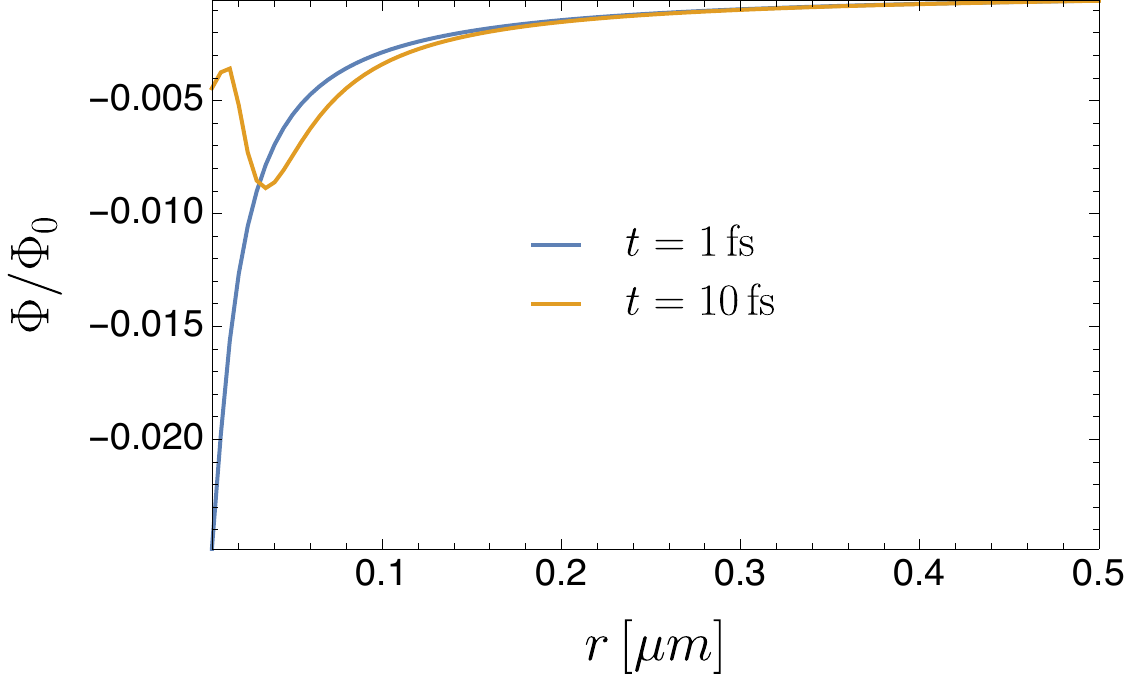} 
\par\end{centering}
\begin{centering}
\includegraphics[width=8cm]{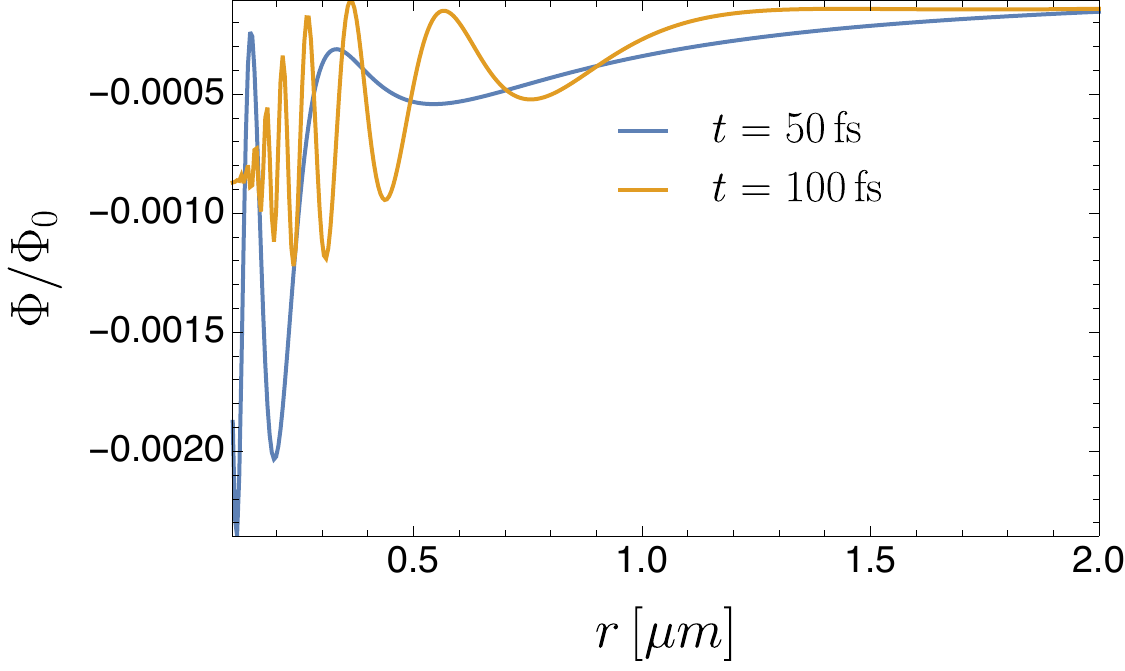} 
\par\end{centering}
\caption{Potential at position $z_{0}=0.1\,{\rm \mu}$m above the graphene
sheet ($E_{F}=0.37$\,eV) for $t=1,\,10$\,fs (top) and $t=50$,
100\,fs (bottom) and for a particle of speed $v=0.01c$. At large
distances the induced potential approaches zero. At shorter times,
we witness the formation of the surface plasmon polariton. At longer
times the disturbance in the electronic density propagates way from
the point $r=0$. The moving electric charge starts at the graphene
sheet. As time evolves oscillatory behavior develops in the induced
potential associated with the modulation of the electronic charge
in the graphene sheet. The particle kinetic energy is about 25\,eV.
Note that the potential oscillations are not characterized by a well
defined wavelength, meaning that surface plasmons of different wave
numbers are excited simultaneously. \label{fig:Potential_perpendicular_motion}}
\end{figure}

One experimental method of accessing graphene surface plasmons is
measuring the energy loss of an electron (or charged particle in general)
when it passes through a graphene sheet. With this in mind we shall
first consider the motion perpendicular to the graphene plane. The
induced potential is given by 
\begin{equation}
\Phi_{{\rm in}}(\mathbf{k},z,\omega)=%Z\frac{n_{0}e^{3}}{4\epsilon_{0}^{2}m_{g}}
\Phi_{0}v^{2}\frac{e^{-k\vert z\vert}}{\omega^{2}-\omega_{{\rm spp}}^{2}}\frac{2vk}{v^{2}k^{2}+\omega^{2}}
\end{equation}
Fourier transforming to real space and time we have 
\begin{multline}
\Phi_{{\rm in}}(\mathbf{r},z,t)=%Zv\frac{n_{0}e^{3}}{4\epsilon_{0}^{2}m_{g}}
\Phi_{0}\\
\times\int v^{2}\frac{d\omega d\mathbf{k}}{(2\text{\ensuremath{\pi}})^{3}}\frac{e^{-k\vert z\vert}e^{i(\mathbf{k}\cdot\mathbf{r}_{\parallel}-\omega t)}}{(\omega+i\eta)^{2}-\omega_{{\rm spp}}^{2}}\frac{2vk}{v^{2}k^{2}+\omega^{2}}
\end{multline}
where $\eta$ is a small positive real number added to account for
causality. The angular integral gives $2\pi J_{0}(kr)$. And after
performing the frequency integral we obtain 
\begin{multline}
\Phi_{{\rm in}}(\mathbf{r},z,t)=%Zv\frac{n_{0}e^{3}}{4\epsilon_{0}^{2}m_{g}}
\Phi_{0}\\
\times\int_{0}^{\infty}v^{3}\frac{k^{2}dk}{2\pi}e^{-k\vert z\vert}J_{0}(kr)I_{1}(\omega_{{\rm spp}},k,t)
\end{multline}
where 
\begin{multline}
I_{1}(\omega_{{\rm spp}},k,t)=-\theta(t)\frac{\sin(\omega_{{\rm spp}}t)}{\omega_{{\rm spp}}}\frac{1}{\omega_{{\rm spp}}^{2}+v^{2}k^{2}}\\
-\frac{1}{2vk}\frac{e^{-vk\vert t\vert}}{v^{2}k^{2}+\omega_{{\rm spp}}^{2}}.
\end{multline}
Therefore the problem of finding the induced electrostatic potential
amounts to a simple quadrature. In figure~\ref{fig:Potential_perpendicular_motion}
we represent $\Phi_{{\rm in}}(r,0,t)$ as function of the distance
to the origin for four different times. For shorter times we see the
formation of the surface plasmon wave. At longer times the surface
plasmon has propagated a given distance. It is clear that the electrostatic
disturbance is not monochromatic since a single wavelength cannot
be identified from the figure. As we will see in the next section
this will translate into an non-trivial spectrum for the energy loss
of a charged particle when it transverses a graphene sheet.

\subsection{The EEL Spectrum}

Here we want to compute the electron-energy loss (EEL) spectrum. This problem has been considered for graphene before using a completely different formalism \cite{Horing-2010}. Using the same methods of the previous reference, the problem of a 2D electron gas has also been considered \cite{Horing-1987,Gumbs-1988,Gumbs-1991}, including the case where  magnetic field is present.

\begin{figure*}
\begin{centering}
\includegraphics[width=14cm]{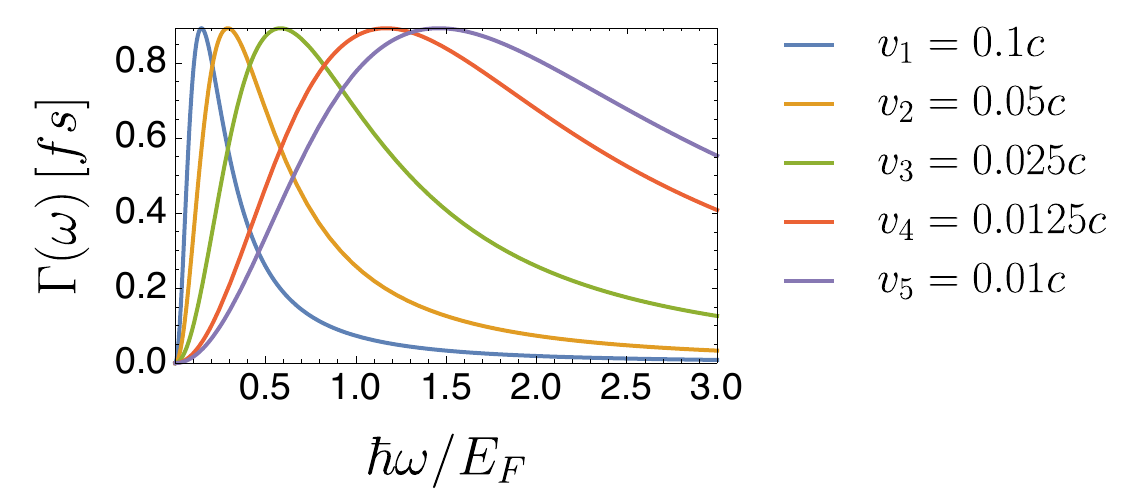} 
\par\end{centering}
\centering{}\caption{Loss spectrum as function of the energy for five speeds to the electron.
The peaks disperse as function\label{fig:Loss-spectrum} of the speed
of the electrons. This allows to retrieve the dispersion of the plasmons.
The wavenumber $k_{{\rm res}}$, which satisfies the condition $\omega_{{\rm spp}}(k_{{\rm res}})=k_{{\rm res}}v$,
is connected to the resonance frequency $\omega_{{\rm res}}$, the
maximum of the EEL spectrum. This relation allows to reconstruct the
surface plasmon dispersion from the EEL spectrum. The Fermi energy
is $E_{F}=0.4$\,eV. The long tail as function of frequency suggests
that a continuum of surface plasmons is excited by the moving charge.}
\end{figure*}

To goal in view we need the quantity $E_{z}(0,z,\omega)=-\partial\Phi_{{\rm in}}(0,z,\omega)/\partial z$
since by definition the EEL spectrum reads~\cite{key-28} 
\begin{equation}
\Gamma(\omega)=\frac{Ze}{\pi\hbar\omega}\int_{-\infty}^{\infty}dt\,\Re \{e^{i\omega t}\mathbf{V}\cdot\mathbf{E}(0,vt,\omega)\}
\end{equation}
where $\mathbf{V}=(0,0,v)$ and $Z=1$ for the electron, and 
the symbol
$\Re$ stands for the real part. The induced
electrostatic potential reads 
\begin{multline}
\Phi_{{\rm in}}(0,z,\omega)=%Zv\frac{n_{0}e^{3}}{4\epsilon_{0}^{2}m_{g}}
\Phi_{0}\\
\times\int_{0}^{\infty}v^{2}\frac{kdk}{2\pi}\frac{e^{-k\vert z\vert}}{(\omega+i\eta)^{2}-\omega_{{\rm spp}}^{2}}\frac{2vk}{v^{2}k^{2}+\omega^{2}}.
\end{multline}
Therefore it follows that the EEL spectrum can be written as

\begin{multline}
\Gamma(\omega)=\frac{Ze}{\pi\hbar\omega}%Zv\frac{n_{0}e^{3}}{4\epsilon_{0}^{2}m_{g}}
\Phi_{0}\int_{-\infty}^{\infty}dt\int_{0}^{\infty}v^{3}\frac{k^{2}dk}{2\pi}\frac{2vk}{v^{2}k^{2}+\omega^{2}}\\
\times\frac{{\rm sign}(t)e^{-k\vert vt\vert}e^{i\omega t}}{(\omega+i\eta)^{2}-\omega_{{\rm spp}}^{2}}
\end{multline}
where the real part is implicit. Performing the time integral we find
(using the Sokhotski\textendash Plemelj theorem) 
\begin{multline}
\Gamma(\omega)=%\frac{Z^{2}v^{2}}{\pi\hbar\omega}\frac{n_{0}e^{4}}{4\epsilon_{0}^{2}m_{g}}
\frac{Ze}{\pi\hbar\omega}\Phi_{0}\int_{0}^{\infty}v^{3}\frac{k^{2}dk}{2\pi}\frac{2vk}{v^{2}k^{2}+\omega^{2}}\\
\times\frac{2\omega}{k^{2}v^{2}+\omega^{2}}\pi\delta(\omega-\omega_{{\rm spp}}).
\end{multline}
Writing $\omega_{{\rm spp}}=\sqrt{ak}$, where the parameter $a$ is:
\begin{equation}
a=2\alpha E_{F}c/\hbar,
\end{equation}
and has units of acceleration {[}see equation~ (\ref{eq:omega_spp}){]},
we can easily integrate the delta function, reading 
\begin{align}
\Gamma(\omega) & =Z^{2}\frac{2\hbar}{E_{F}}\frac{\omega^{2}v^{2}/a^{2}}{\left(\omega^{2}v^{2}/a^{2}+1\right)^{2}}\label{eq:EELs_simple}
\end{align}
a result that has been obtained in the literature before~\cite{key-18}
using a different method based on reflection coefficients. Equation~(\ref{eq:EELs_simple})
has a maximum at the frequency 
\begin{equation}
\hbar\omega_{{\rm res}}/E_{F}=2\alpha\frac{c}{v}\label{eq:omega_res}
\end{equation}
corresponding to an efficient excitation of surface plasmons of that
frequency. We plot $\Gamma(\omega)$ in figure~\ref{fig:Loss-spectrum}.
From this figure we see the dispersion shifts towards higher energies
as the speed of the moving electron decreases, in agreement with equation~(\ref{eq:omega_res}).
Looking at the frequency where the EEL spectrum has a maximum we can
find the surface plasmon frequency. This frequency coincides with
the interception of the $\omega_{{\rm spp}}$ curve with the line
$\omega=k_{z}v$. This allows to retrieve the wavenumber $k_{z}$
of the surface plasmon associated with the $\omega_{{\rm res}}$ obtained
from the EEL spectrum. We note, however, that this process of exciting
surface plasmons does not produce a monochromatic wave, as can be
guesses from the broadening of the EEL spectrum and from figure~
\ref{fig:Potential_perpendicular_motion}. Note that from the latter
figure we cannot attribute a single wavelength to the potential disturbance.
As we will see, the motion of an electron parallel to a graphene sheet
is able to induce a monochromatic plasmon.

%\begin{widetext}

%\end{widetext}

\subsection{Motion parallel to the graphene sheet: Kelvin and Mach wakes}

\begin{figure}[h!]
	\begin{centering}
		\includegraphics[width=8cm]{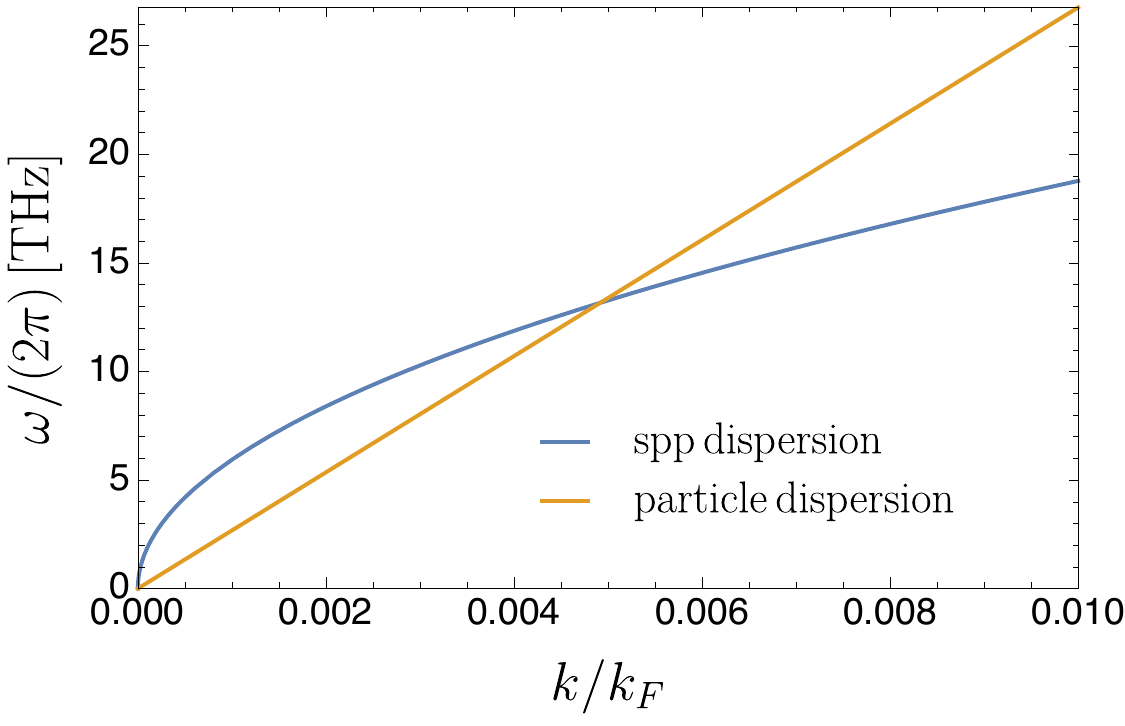} 
		\par\end{centering}
	\begin{centering}
		\includegraphics[width=8cm]{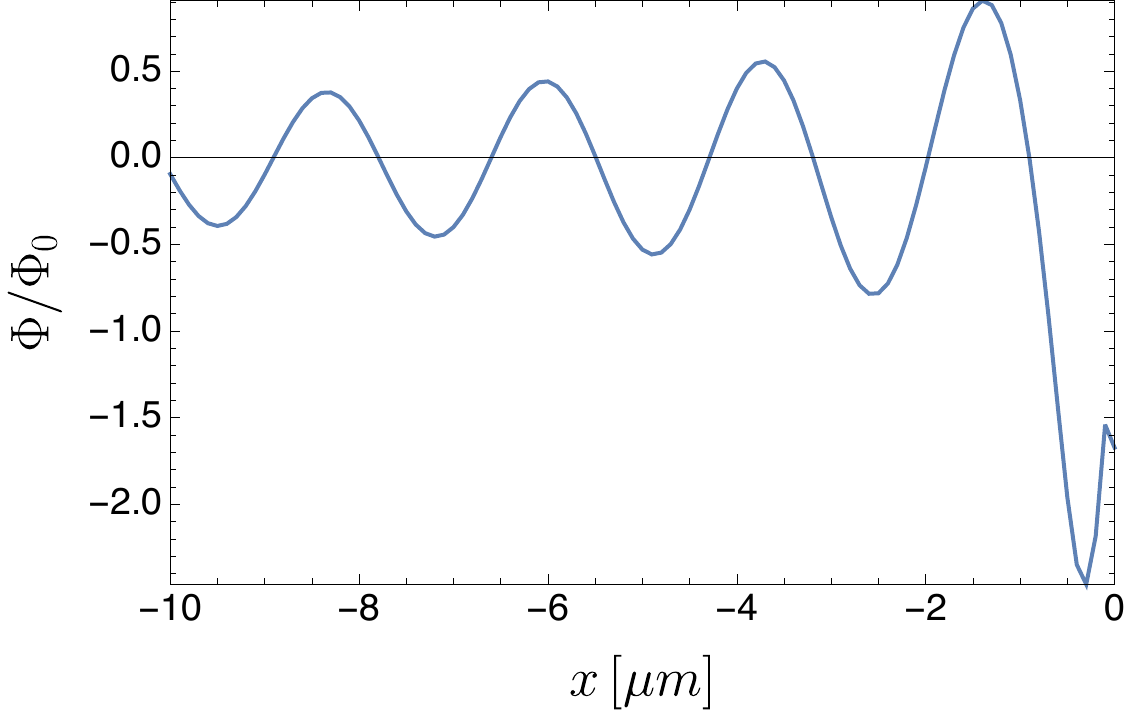} 
		\par\end{centering}
	\caption{Spectrum (top) and electrostatic potential (bottom) along the direction
		of propagation of the moving particle. The particle dispersion, given
		by $\omega=kv$, is represented on the left image by the orange line.
		It intercepts the SPP dispersion at $k/k_{F}=0.00495$, corresponding
		to a SPP wavelength of $\lambda_{{\rm spp}}=2\pi/k\approx2.3\,{\rm \mu}$m.
		This value matches well the distance between successive crests of
		the potential, $\Delta y\approx2.4\,{\rm \mu}$m. Changing the velocity
		of the particle changes the slope of the orange line and therefore
		the wavelength of the SPP, which turns into different distances between
		two successive crests of the potential. The parameters are: $E_{F}=0.37$\,eV,
		corresponding to an electronic density equal to $10^{5}\,{\rm \mu}$m$^{-2}$,
		$v=0.1c$, $z_{0}=0.01\,{\rm \mu}$m, and $t=0$\,fs.}
\end{figure}

We now consider the case of a charge moving parallel to a graphene
sheet at a constant height $z_{0}$ (experimentally we could consider
EEL with a tilted graphene sample). The calculation of the potential
when the charge moves parallel to graphene follows from the Fourier
representation of the charge density. In this case the calculation
is simpler due to the presence of two Dirac delta-functions, but an
integral in the coordinates $k_{x}$ and $k_{y}$ still persists.
Let us compute the potential $\Phi_{{\rm in}}(\mathbf{r},z,t)$ starting
from
\begin{equation}
\Phi_{{\rm in}}(\mathbf{k},z,\omega)=2\pi%Z\frac{n_{0}e^{3}}{4\epsilon_{0}^{2}m_{g}}
\Phi_{0}v^{2}\frac{e^{-k(\vert z\vert+z_{0})}}{\omega^{2}-\omega_{{\rm spp}}^{2}}\delta(\omega-vk_{y}).\label{eq:potential_longitudinal_motion}
\end{equation}
Note that the delta-function implies that the particle disperses with
a frequency given by $\omega=vk_{y}$.

Fourier transforming equation~(\ref{eq:potential_longitudinal_motion})
to real space and time, we obtain in polar coordinates

\begin{align}
\Phi_{{\rm in}}(\mathbf{r}_{\parallel},z,t)= & \Phi_{0}v^{2}\int_{0}^{2\pi}\frac{d\theta}{2\pi}\int_{0}^{\infty}\frac{kdk}{2\pi}\nonumber \\
\times & \frac{e^{-k(\vert z\vert+z_{0})}e^{ik[r\cos(\theta-\theta^{\prime})-vt\cos\theta]}}{(k\cos\theta v)^{2}+i{\rm \,sign}(\cos\theta)\eta-ak}
\end{align}
where $\theta^{\prime}$ is the polar angle of $\mathbf{r}_{\parallel}$.
Using the Sokhotski\textendash Plemelj theorem it follows that ($\fint$
stands for principal value of the integral) 
\begin{multline}
\Phi_{{\rm in}}(\mathbf{r}_{\parallel},z,t)=\Phi_{0}\int_{0}^{2\pi}\frac{d\theta}{2\pi}\fint_{0}^{\infty}\frac{dk}{2\pi}\frac{e^{-k\beta}e^{ik\gamma}}{k\cos^{2}\theta-a/v^{2}}\\
-i\pi\Phi_{0}\int_{0}^{2\pi}\frac{d\theta}{2\pi}\int_{0}^{\infty}\frac{dk}{2\pi}e^{-k\beta}e^{ik\gamma}{\rm sign}(\cos\theta)\\
\times\delta(k\cos^{2}\theta-a/v^{2})
\end{multline}
Let us introduce the change of variables $k\cos^{2}\theta-a/v^{2}=\kappa$
which modifies the integral to 
\begin{align}
\Phi_{{\rm in}}(\mathbf{r}_{\parallel},z,t)= & \Phi_{0}\int_{0}^{2\pi}\frac{d\theta}{\left(2\pi\right)^{2}}\frac{e^{-f(\theta)}}{\cos^{2}\theta}\fint_{-a/v^{2}}^{\infty}d\kappa\frac{e^{-v^{2}f(\theta)\kappa/a}}{\kappa}\nonumber \\
- & i\pi\Phi_{0}\int_{0}^{2\pi}\frac{d\theta}{\left(2\pi\right)^{2}}\frac{e^{-f(\theta)}}{\cos^{2}\theta}{\rm sign}(\cos\theta)
\end{align}
where $\beta=\vert z\vert+z_{0}$, $\gamma=[r\cos(\theta-\theta^{\prime})-vt\cos\theta]$,
and 
\begin{equation}
f(\theta)=\frac{a}{\cos^{2}\theta v^{2}}\left[\beta-i\gamma(\theta)\right].
\end{equation}
The principal value of the integral over $\kappa$ gives the exponential
integral function, ${\rm {\rm Ei}}(x)$, and the integral of the delta
function is elementary. It then follows: 
\begin{align}
\Phi_{{\rm in}}(\mathbf{r}_{\parallel},z,t)= & -\Phi_{0}\int_{0}^{2\pi}\frac{d\theta}{(2\pi)^2}\frac{e^{-f(\theta)}}{\cos^{2}\theta}{\rm {\rm Ei}}[f(\theta)]\nonumber\\
- & i\pi\Phi_{0}\int_{0}^{2\pi}\frac{d\theta}{(2\pi)^2}\frac{e^{-f(\theta)}}{\cos^{2}\theta}{\rm sign}(\cos\theta).
\label{eq:plasmonic_wake}
\end{align}

\begin{figure}[t!]
	\begin{centering}
		\includegraphics[width=8cm]{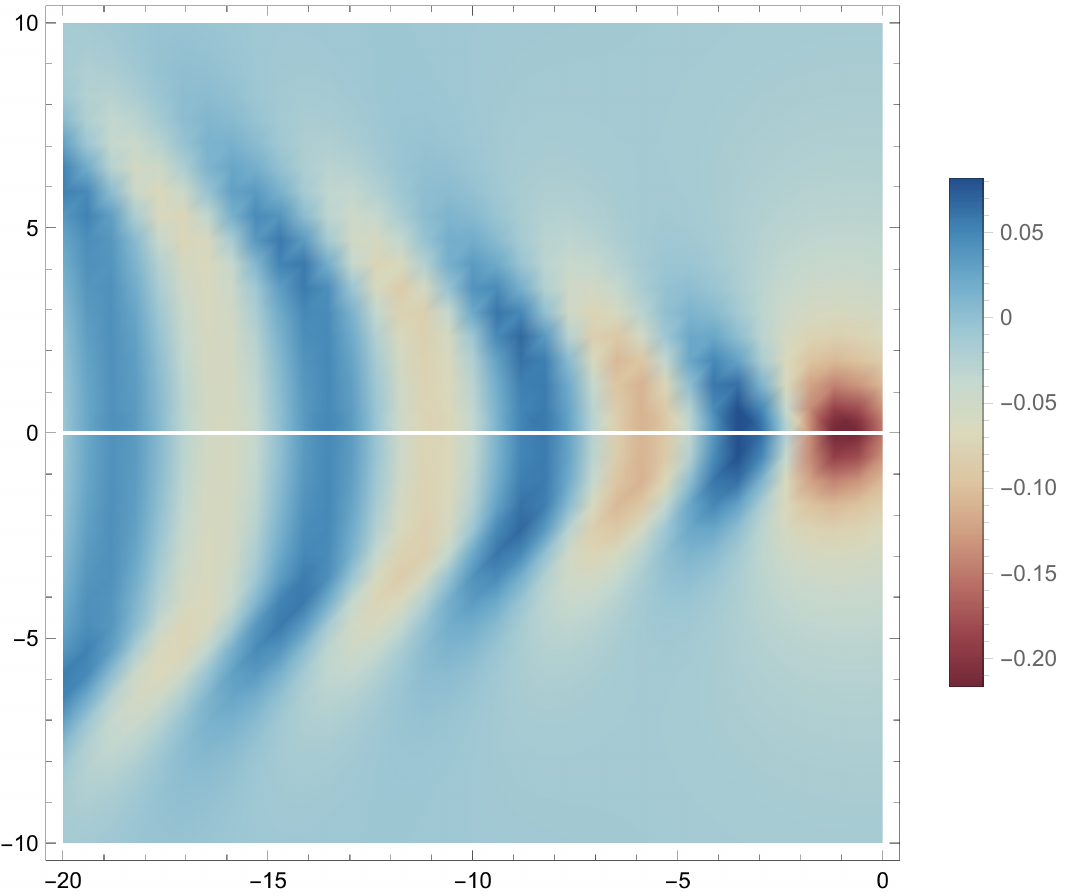} 
		\par\end{centering}
	\centering{}\centering{}\includegraphics[width=8cm]{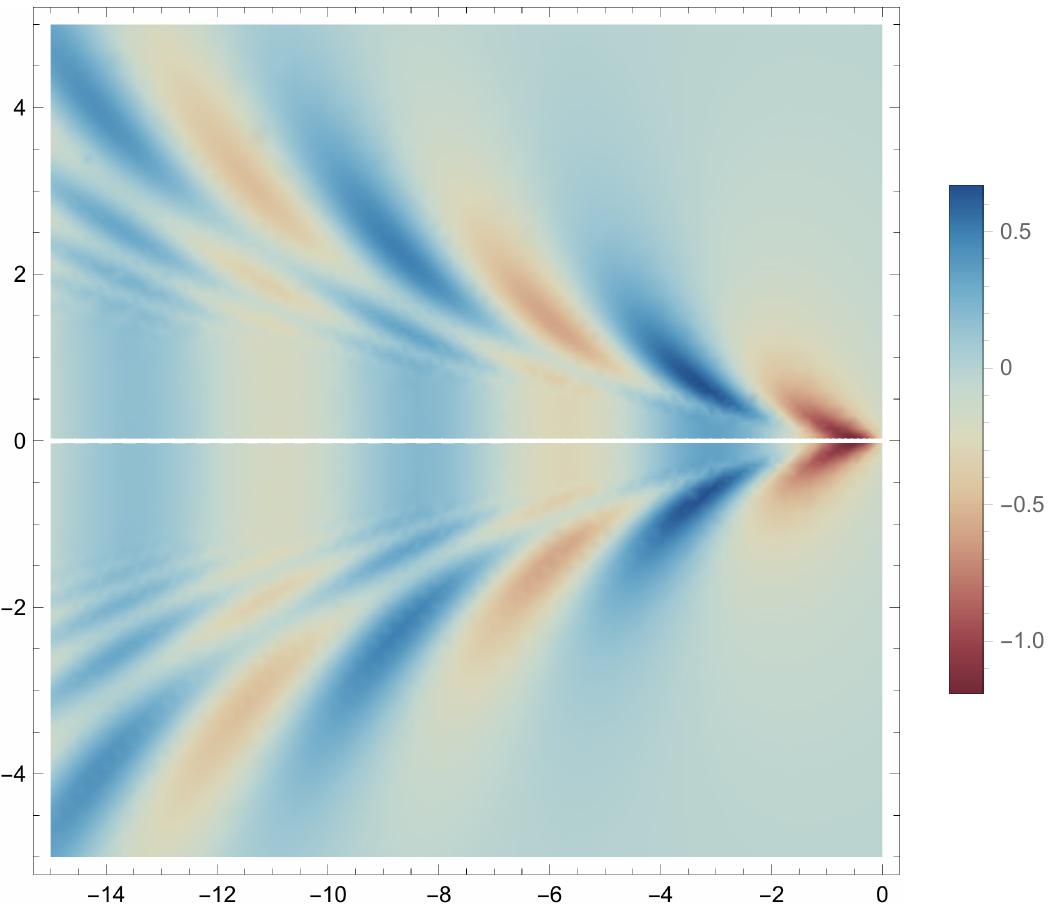}
	\caption{Electrostatic potential, in units of $\Phi_0$, in graphene for a particle moving parallel
		to it at $z_{0}=1\,{\rm \mu}$m with speed $v=0.1c$ (top panel; $E_{F}=0.17$\,eV)
		and at $z_{0}=0.1\,{\rm \mu}$m with speed $v=0.1c$ (bottom panel;
		$E_{F}=0.17$\,eV). 
		The axes of the figures are in $\mu$m.
		A Kelvin wake develops on the graphene sheet
		in the image of the top panel. Indeed, in the top panel the half-opening
		angle is about $\sim(22.8\pm0.3){\rm o}$, whereas in the bottom one is
		about $\sim(16.8\pm 0.3)^{{\rm o}}$. Note that the electrostatic potential fluctuations are much larger for the case depicted in the bottom panel, due to a closer proximity of the moving charge to graphene. 
		Also note the presence of a plane wave in the central region of the wake (more evident in the bottom panel).
		(The horizontal and vertical scales in the two panels are different.)
		\label{fig:Kelvin}}
\end{figure}

In figure~\ref{fig:Kelvin} we show two examples of the electrostatic
potential induced by the charged particle moving relatively to graphene
at a distance $z_{0}$ from it. It is evident that the surface plasmons
propagate in the form of ship wakes. Contrary to conventional wisdom,
ship wakes are not~\cite{key-4} necessary given by Kelvin theory
and the same happens for plasmonic wakes induced by the moving charge.

According to Kelvin theory half the angle of the cone of a ship
wake is given by 
\begin{equation}
\theta_{K}=\arctan\frac{1}{\sqrt{8}}\approx19.47^{{\rm o}}\label{eq:Kelvin}
\end{equation}
that is, it is a constant number independent of the speed of the ship.  For future notice, it
is important to clarify how the opening angle of the plasmon wake is
determined from our simulations: for a fixed value of $\vert\mathbf{r}\vert$ we  compute
from the electrostatic potential the three components of the electric
field as function of the angle $\theta^{\prime}$; we then use these
results to compute the absolute value of the electric field and fit a Gaussian of the form $Ae^{-(\theta^\prime-\theta_{\rm max})^2/\sigma}$,
where $\theta_{\rm max}$ is the angle where the intensity of the electric field is maximum, and $A$ and $\sigma$
are fitting parameters;
the
opening half angle is defined as the angle where the fitting function has the value of 0.61 of its maximum, at a radial position value given by 
$\vert\mathbf{r}\vert=(2+1/4)2\pi v^2/a$ (other choices of $\vert\mathbf{r}\vert$ lead to the same results). Using this procedure and with the help of dimensional
analysis we have found that half the aperture of the cone is given
with good accuracy by (in degree)
\begin{equation}
\theta^\prime\approx\frac{1}{\delta}\sqrt{\frac{z_{0}a}{v^{2}}}
\equiv\frac{1}{\delta}\frac{1}{{\rm Fr}_{\rm pl}}
\label{eq:theta_guess}
\end{equation}
for $\rm {Fr}_{pl}\gg1$, where $\delta$ is a real constant
that we have found to be of the order of $\delta\approx0.019\pm0.001$ 
($\delta\approx 1.09$ for $\theta^\prime$ in radians)
and ${\rm Fr}_{\rm pl}$ is the plasmonic Froude number (see discussion in  section \ref{sec:Conclusions}). We note that formula
(\ref{eq:theta_guess}) should work well only in the large
Froude number regime and should be understood as the first term in powers of $1/{\rm Fr}_{\rm pl}$ of a more complex expression.
From the numerical data,
we have
identified, a transition from a Mach-type wavefront, where the opening
angle of the plasmon wake follows the law $\theta\propto1/v$, at
high speeds (for the remaining parameters fixed) to a 
Kelvin-type
one, where the angle of the plasmon wake is independent of the speed
of the moving charge (see discussion ahead). Indeed, for slow speeds compared to $c$ (and small Froude numbers) the
wavefront is always Kelvin-like, that is, with an opening angle for
the wake independent of the speed of the moving charge. For the parameters
of the top panel of figure~\ref{fig:Kelvin} formula~(\ref{eq:theta_guess})
predicts an angle of 56$^{{\rm o}}$ well above Kelvin's value,
whereas from the figure we estimate a value of about $\sim22.8{}^{{\rm o}}$.
However, we note that the prediction is outside the validity of the
condition $\rm {Fr}_{pl}\gg1$ and therefore quantitative disagreement
is expected. In this case the Froude number reads $\rm {Fr}_{pl}=0.9$. For the bottom panel of the same figure, formula~(\ref{eq:theta_guess})
predicts a value of $(18\pm3)^{{\rm o}}$, whereas from the electric field
intensity we estimate a value of $\sim16.8{}^{{\rm o}}$, which is
in very good agreement with the result given by equation (\ref{eq:theta_guess}). Note that in this case the Froude number is $\rm {Fr}_{pl}=2.86$.
Also, note that from the top to the bottom panel, $z_{0}$ has changed by
one order of magnitude. 

\begin{figure}
	\centering{}\centering{}\includegraphics[width=8cm]{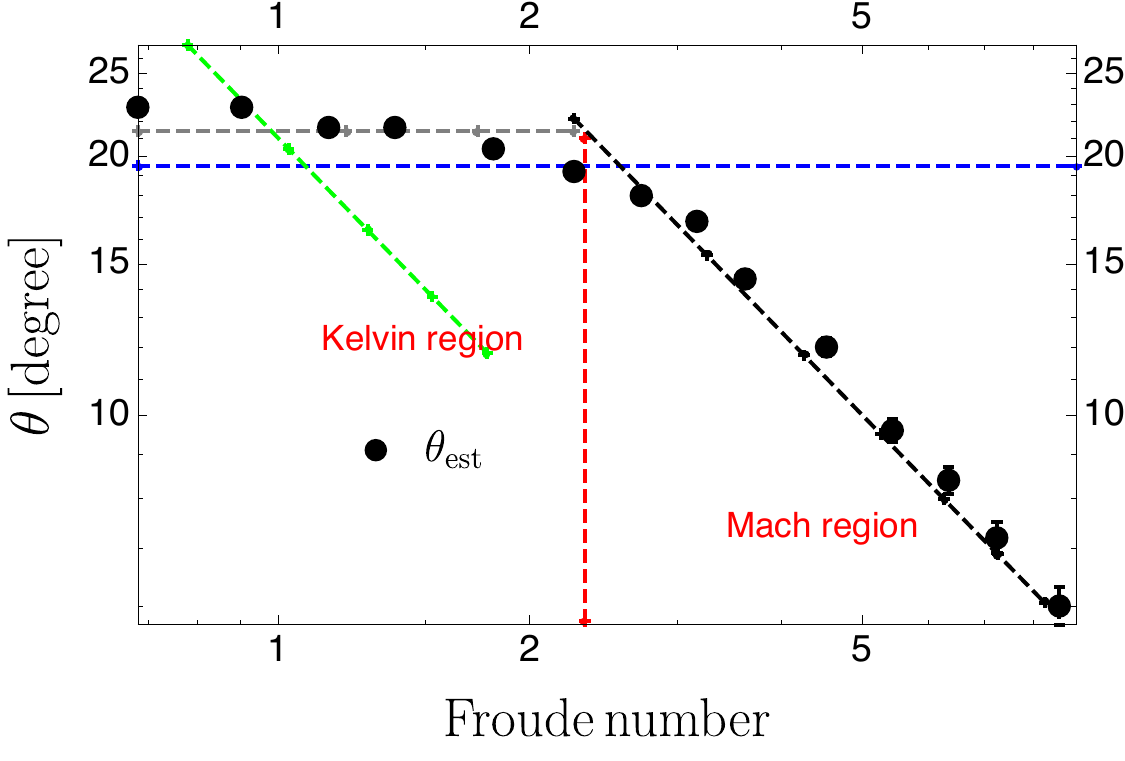}
	\includegraphics[width=8cm]{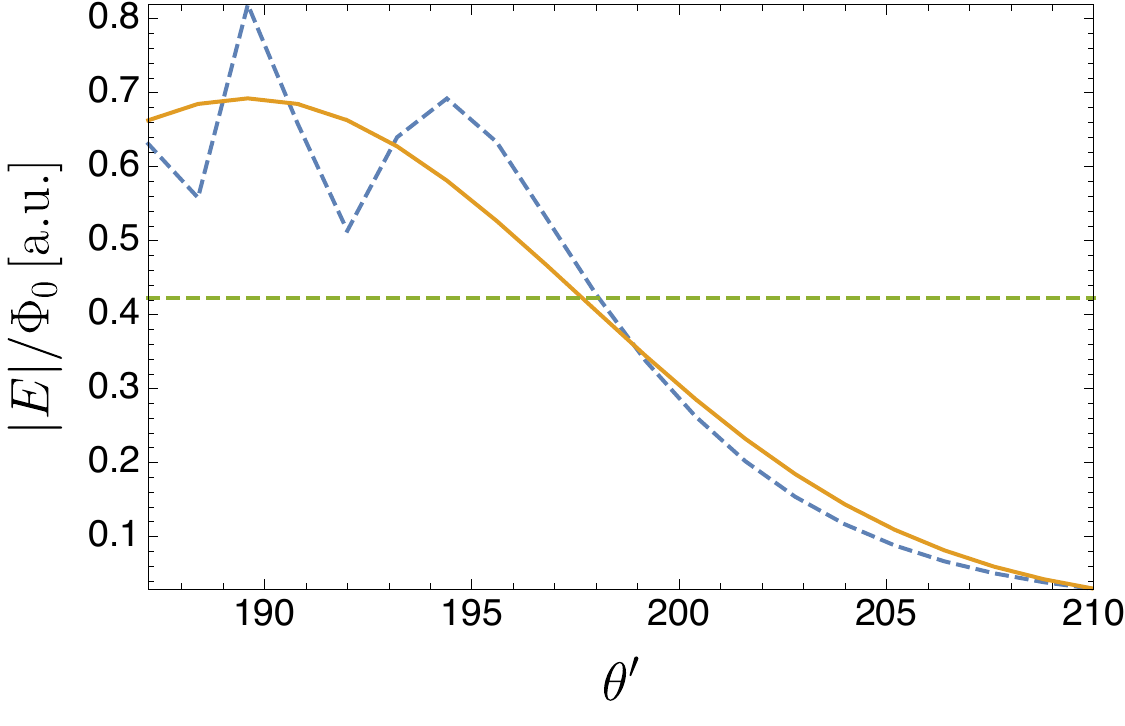}
	\caption{Graphical representation of the data of table \ref{tab:Estimation-of-half-angle}. Top panel:
		Note the transition at the Froude number $\rm{Fr_{pl}}\sim2$ from Mach-type to Kelvin-type
		of wake, as the Froude number decreases. The parameters are $z_{0}=1\,{\rm \mu}$m and $E_{F}=0.17$\,eV, and the angle was measured at a distance $\vert {\mathbf r}\vert=(2+1/4)2\pi v^2/a$ (corresponding to two wavelengths plus one quarter) from the apex of the cone. The error of the data points was estimated to be
		0.3$^{\rm o}$.
		Different parameters will give similar curves to this one. The angle
		$\theta_{{\rm est}}$ is estimated from the electric field, for a
		given $v/c$ ratio, as explained in the text. The dashed black line
		represents the  angle given by the formula $\theta=1/(\delta{\rm Fr}_{\rm pl} )$  as function of the Froude number (this formula only holds in the Mach-type region of the Froude number). The horizontal dashed
		blue line represents Kelvin's result  (see section \ref{sec:Conclusions} for a more thorough discussion of the role of the Froude number).
		The horizontal gray line is a fit to the simulation data points in the Kelvin region  (the transition from Mach-like to Kelvin-like behavior is clearly seen to occur at the interception of the gray and black dashed curves). The green dashed line is an attempt to fit the simulation data points in Kelvin's region with the same expression used in the Mach region (this clearly fails to account for the behavior of the data).   Bottom panel: here we give an example of the fitting procedure (see text for details). The dashed blue line is the simulation data, the orange solid line is the 
		fit of the data to a Gaussian, and the dashed green line signals the value of 0.61 of the maximum of the fitting function which  approximately corresponds to the inflection point of the curve ($0.61\approx e^{-1/2}$). The interception of the 
		horizontal line with the solid curve defines the half angle of aperture of the wake cone. The speed of the particle is $v=0.1c$, $E_F=0.17$\,eV,  and $z_0=0.1\,\mu$m, corresponding to a Froude number of 
		$\rm{Fr_{pl}}=$2.86. In this example we have
		$\theta_{\rm est}\approx196.8-180=16.8^{\rm o}$.
		 \label{fig:transition}}
\end{figure}

A study of the evolution of the plasmon wake
from Kelvin-type to Mach-type is given in figure~\ref{fig:transition}.
Note the transition located at $\rm Fr_{pl}\sim2$ from a Mach-type wake
to a Kelvin-type one, as the Froude number decreases. The existence of such transition was first pointed
out by Shi~\emph{et al.}~\cite{key-3}, who solved an identical
problem numerically but gave no interpretation to the phenomenon as they were unable to identify the Froude number for graphene. A study of the prediction given by equation~(\ref{eq:theta_guess})
and the estimation based on the Figures is given in table~\ref{tab:Estimation-of-half-angle}.
The agreement between the numbers in the two rows is good for values
of $\rm Fr_{pl}$ larger than 2, showing that the ansatz 
$\theta=1/(\delta{\rm Fr}_{\rm pl})$
does a good job at predicting the values obtained from the calculation
of the absolute value of the electric field. 
The last row gives the Froude number. We have, therefore, gathered evidence for the existence of  a transition from 
Mach-type to Kelvin-type waves at a critical Froude number
of $\rm {Fr}_{pl}^c\gtrsim2$ (note that the ratio $v/c$ is not the good quantity to analyze this problem).

\begin{table*}
\begin{centering}
\begin{tabular}{|c||c|c|c|c|c|c|c|c|c|c|c|c|c|c|}
\hline 
$v/c$   & 0.075   & 0.1   & 0.125&0.15  & 0.2  & 0.25  & 0.3  & 0.35& 0.4  & 0.5  & 0.6  & 0.7  & 0.8  & 0.95\tabularnewline
\hline 
\hline 
$\theta_{{\rm est}}$  & 22.8    & 22.8  & 21.6  & 21.6  & 20.4  & 19.2  & 18   & 16.8 & 14.4  & 12  & 9.6  & 8.4  & 7.2  & 6\tabularnewline
\hline 
$\theta^\prime$   & 74 & 56 & 44 & 37 & 28  & 22.2  & 18.5  & 15.9  & 13.9  & 11.1  & 9.3    & 7.9  & 6.9  & 5.8\tabularnewline
\hline
${\rm Fr}_{\rm pl}$ & 0.68 &0.91& 1.1 & 1.4 & 1.8 & 2.3 &2.7 & 3.2&3.6 &
4.5 & 5.4 & 6.3 & 7.2 & 8.6
\tabularnewline
\hline 
\end{tabular}
\par\end{centering}
\caption{Estimation of half the angle, $\theta_{{\rm est}}$, of the cone associated
with the moving plasmons induced by a moving charge compared with
the prediction of equation $\theta^\prime=1/(\delta{\rm Fr}_{\rm pl})$, with $\delta\approx0.019$. Note that the validity of this formula is restricted to  
${\rm Fr}_{\rm pl}>2$. The parameters
are $E_{F}=0.17$\,eV and $z_{0}=1\,{\rm \mu}$m. Figure~\ref{fig:transition}
plots the numbers of the table in a log-log scale. Also note the transition
at about $v/c\approx0.25$. (See text for the method used to estimate
$\theta_{{\rm est}}$.) 
The last line gives the plasmonic Froude number
(see discussion in  section \ref{sec:Conclusions}), 
${\rm Fr}_{\rm pl}=\sqrt{v^2/(z_0a)}$.
It is clear that the criterion for the validity of formula 
(\ref{eq:theta_guess}) is related to the value of the Froude number: When ${\rm Fr}_{\rm pl}\gtrsim 2$ the formula works well.
\label{tab:Estimation-of-half-angle}}
\end{table*}

For the parameters considered in figure \ref{fig:transition},
and as noted above, we have $\rm Fr_{pl}\approx2.3$  for $v/c=0.25$, the speed at which the transition from the Mach-like to Kelvin-like regime occurs (for the given parameters). Therefore the transition between the two regimes is controlled by Froude number, with the transition occurring roughly for $\rm Fr_{pl}\approx2$. This result should be a generic feature of plasmonic wakes in graphene. 
Finally, we note that the region of Kelvin-type has a constant angle (gray dashed line in figure \ref{fig:transition}) of $(21.4\pm 1.4)^{\rm o}$, a value larger than that predicted by Kelvin's theory, but with Kelvin's value within the interval of uncertainty. This larger value of the opening angle happens due to the definition we have used 
for determining it. Indeed, adopting a 
slightly different criterion the angle values would be slightly different, but the transition would occur at the same Froude number (results not shown). That is, the existence of a transition from Mach-like to 
Kelvin-like behavior does not depend on the criterion used
to define the opening angle of the cone.
We shall discuss more on all the above in section \ref{sec:approx} and in
section \ref{sec:Conclusions}.

We can also compute the power added to the charged particle using the expression
$
{\cal P}=  Ze\mathbf{V}\cdot\mathbf{E}(0,vt,z_{0},t),
$
where $\mathbf{V}=(0,v,0)$ and $\mathbf{E}=\mathbf{E}(x,y,z,t)$.
This means that we need to compute the $y$-component of the field.
This follows from $E_{y}=-\partial\Phi_{{\rm in}}(\mathbf{r_{\parallel}},z,t)/\partial y$.
Once the electric field is known, the power  can
be determined. The calculation is consistent if the energy loss is
small compared to the kinetic energy of the particle. Since the field
$E_{y}(0,vt,z_{0},t)$ is time independent the power loss is also time
independent. We have 
verified in our numerics (results not shown)  the energy loss is of the order of 2\,meV,
for a particle with speed $v=0.1c$ (kinetic energy 2.5\,keV) propagating
over graphene a distance of $3\,{\rm \mu}$m). We note that the situation is different
for the motion of a charged particle perpendicular to the graphene
plane due to lack of translation invariance. In this latter case
the power loss is time dependent.

Although numerical calculations are a powerful way of gaining understanding about  a complex problem, much insight can be gained
from deriving analytical results, even when they are only valid in special limits. In the present section we have conducted a fully 
numerical analysis of the plasmonic wake. In the next section two special limits are considered where it became possible to obtain
closed analytical expressions for the shape of the wake.

\subsection{Approximate analytical formulas for the plasmonic wake \label{sec:approx}}

In this section we derive analytical formulas for the plasmonic wakes valid in the limits $az_0/v^2\gg1$ and $az_0/v^2\ll1$.
The first limit corresponds approximately  to the case of the top panel of 
figure \ref{fig:Kelvin} ($az_0/v^2\simeq1.2$) whereas the second limit corresponds to the 
bottom panel of the same figure ($az_0/v^2\simeq0.12$). 
The exact form of the plasmonic wake is given by equation (\ref{eq:plasmonic_wake}). However, we are interested here in obtaining approximate analytical expressions for the wake, which can then be used to gain some insight on its properties.  To this end, we consider the asymptotic expression for the Ei$(z)$ function, which to leading order reads ${\rm Ei}(z)\sim e^z/z$ ($\Re z>0$). This shows that to leading order the first integral in equation (\ref{eq:plasmonic_wake}) is elementary and does not contribute significantly to the form of the wake as this comes from the exponential in the second integral [a careful numerical study of both integrals in equation (\ref{eq:plasmonic_wake}) shows that this statement is approximately correct in some regimes].  In order to derive the needed asymptotic expressions we note that we need to compute the real part of the following integral [ignoring, for the time being,  contributions coming from
the first integral in equation (\ref{eq:plasmonic_wake})]:

\begin{equation}
I_2=-i\pi\int_{0}^{2\pi}d\theta\frac{e^{-\frac{az_0}{v^2\cos^2\theta}}e^{\frac{iar\cos(\theta-\theta^\prime)}{v^2\cos^2\theta}}}{\cos^2\theta}{\rm sign}(\cos\theta)
\end{equation}
which can be shown to equal
\begin{align}
\Re I_2&=2\pi \int_{-\pi/2}^{\pi/2}d\theta\frac{e^{-\frac{az_0}{v^2\cos^2\theta}}}{\cos^2\theta}\sin\left(
\frac{ar}{v^2}\frac{\cos\theta^\prime}{\cos\theta}
\right)\times 
\nonumber\\
&\cos\left(
\frac{ar}{v^2}\frac{\tan\theta}{\cos\theta}\sin\theta^\prime
\right)
\end{align}
We now introduce the change of variable 
$u=\tan\theta$. This implies $ 1/\cos^2\theta=1+u^2$
and $du=\sec^2\theta d\theta$.
Therefore the integral reads

\begin{align}
\Re I_2 &=2\pi e^{-az_0/v^2}\int\limits_{-\infty}^{\infty}du
e^{-u^2az_0/v^2}
\sin\left(
\frac{ar}{v^2}\cos\theta^\prime\sqrt{u^2+1}
\right)
\nonumber\\
&\times
\cos\left(
\frac{ar}{v^2}\sin\theta^\prime u\sqrt{u^2+1}
\right)	
\label{eq:ReI}
\end{align}
Next we observe that for $az_0/v^2\gg1$ the kernel of the integral is strongly peaked at $u=0$, due to the Gaussian exponential. Therefore, in this regime, we introduce the approximation

\begin{align}
\Re I_2&\approx 
2\pi e^{-az_0/v^2}\int\limits_{-\infty}^{\infty}du
e^{-u^2az_0/v^2}
\cos\left(
\frac{ar}{v^2}\sin\theta^\prime u
\right)	
\nonumber\\
&\times
\sin\left(
\frac{ar}{v^2}\cos\theta^\prime(1+u^2/2)
\right)	
\label{eq:large_approx}
\end{align}
which, applying the exponential representation of the trigonometric formulas, can be seen as a Gaussian integral, which has the elementary solution:
%\begin{widetext}
\begin{equation}
\Re I_2\approx -4\pi^{3/2} e^{-\beta}
\frac{ \Im\left(\sqrt{2 \beta -i \gamma } e^{-i \gamma-\frac{\lambda ^2}{4 \beta +2 i \gamma }}\right)}{\sqrt{8 \beta ^2+2 \gamma ^2}}
\label{eq:large_alpha}
\end{equation}
%\end{widetext}
where
\begin{subequations}
\begin{eqnarray}
\beta &=& \frac{az_0}{v^2}\\
\gamma&=&\frac{ar}{v^2}\cos\theta^\prime\\
\lambda&=&\frac{ar}{v^2}\sin\theta^\prime
\end{eqnarray}
\end{subequations}
For obtaining the second term in equation (\ref{eq:plasmonic_wake}) we have to divide equation (\ref{eq:large_alpha}) by $4\pi^2$.
The obtained expression is valid for arbitrary large values of $r$ and describes qualitatively the formation of the wake due to the moving charge in its regime of validity. 
As noted above, the first integral in equation (\ref{eq:plasmonic_wake}) contributes little to leading order in the form of the asymptotic expression of the exponential integral
function. However  the regime $az_0/v^2\gg1$ is likely to be experimentally challenging to access (see discussion below). Therefore, we would like to have an equation
holding in the regime $az_0/v^2\gtrsim1$. Fortunately, this can be obtained treating the first integral approximately.
The procedure is similar to that described above, except that in the end we still have to evaluate the additional integral coming from the 
principal value of the integral in the variable $\kappa$.
The final result to the first integral in equation (\ref{eq:plasmonic_wake}) reads
(up to lowest order in the expansion of the arguments of the trigonometric functions)

\begin{equation}
I_1\approx
\Re\left[
\frac{ \sqrt{2}   (-2 \beta +i \gamma )}{\pi\sqrt{(2 \beta -i \gamma )^2 (2 \beta +i \gamma )}}
D(s)
\right]
\label{eq:I1_small_Froude}
\end{equation}
where $s=\sqrt{i\phi+\psi}$  and $D(z)$ is the Dawson integral, $D(z)=e^{-z^2}\int_0^ze^{y^2}dy$, and
\begin{subequations}
\begin{align}
\psi&=\beta+\frac{4 \beta  \lambda ^2}{16 \beta ^2+4 \gamma ^2}\,,\\
\phi&=\frac{\gamma  \left(8 \beta ^2+2 \gamma ^2-\lambda ^2\right)}{2 \left(4 \beta ^2+\gamma ^2\right)}\,.	
\end{align}	
\end{subequations}
Therefore,
the electrostatic potential  (\ref{eq:plasmonic_wake}) is approximately given by the sum: $I_1+(2\pi)^{-2}\Re I_2$. 
This result is in quantitative agreement to the exact fully numerical calculation of equation 
(\ref{eq:plasmonic_wake}). A better analytical approximation 
than
equation (\ref{eq:I1_small_Froude}) 
to the first integral in equation (\ref{eq:plasmonic_wake})
can be 
obtained, but the resulting expression is too cumbersome to be given here.

Let us next consider the regime $az_0/v^2\ll1$. In this case the integral is dominated by  values of $u$ in a large range centered at  $u=0$. Therefore, the arguments of the trigonometric functions  
are expanded differently than before as:
\begin{align}
\Re I_2&\approx 
2\pi e^{-az_0/v^2}\int\limits_{-\infty}^{\infty}du
e^{-u^2az_0/v^2}
\sin\left(
\frac{ar}{v^2}\cos\theta^\prime\vert u\vert 
\right)
\nonumber\\
&\times\cos\left(
\frac{ar}{v^2}\sin\theta^\prime (u^2+1/2)
\right)		
\end{align}
The integral can be expressed in terms of the Error function, erf$(x)$, as
%\begin{widetext}
\begin{align}
\Re I_2&\approx -4\pi^{3/2} e^{-\beta}
  e^{-\frac{\beta  \gamma ^2}{2 \left(\beta ^2+\lambda ^2\right)}}
  \nonumber\\
  &\times \Re\left[\frac{e^{\frac{\gamma ^2}{4 \beta +4 i \lambda }+i\lambda/2} \rm{erf}\left(\frac{\gamma }{2 \sqrt{-\beta +i \lambda }}\right)}{\sqrt{-\beta +i \lambda }}\right]
\label{eq:small_alpha}
\end{align}
%\end{widetext}
Again, the previous expression is valid for an arbitrary large $r$.
It is interesting to note that the Error function often appears in diffusion problems. Whether the propagation of the surface plasmons in this regime can be seen as a diffusion problem requires more work. Also in this case, we can obtain an expression for the first integral in equation (\ref{eq:plasmonic_wake}). Proceeding as briefly described 
in the regime $az_0/v^2\gg1$, the
 expression for $I_1$ reads
\begin{equation}
I_1\approx -\frac{\Re\left[
	\sqrt{\beta-i\lambda} e^{-s}{\rm erfi}(s)
	\right]}{\pi^{1/2}\sqrt{\beta^2+\lambda^2}}
\label{eq:I1_large_Froude}
\end{equation}
where in this case $s=(4\beta^2+\gamma^2+6i\beta\lambda-2\lambda^2
)/(4\beta+4i\lambda)$ and ${\rm erfi}(z)$ is the complex error function. 
As before, the electrostatic potential (\ref{eq:plasmonic_wake}) is approximately given by the sum: $I_1+(2\pi)^{-2}\Re I_2$.

\begin{figure}[t!]
	\begin{centering}
		\includegraphics[width=8cm]{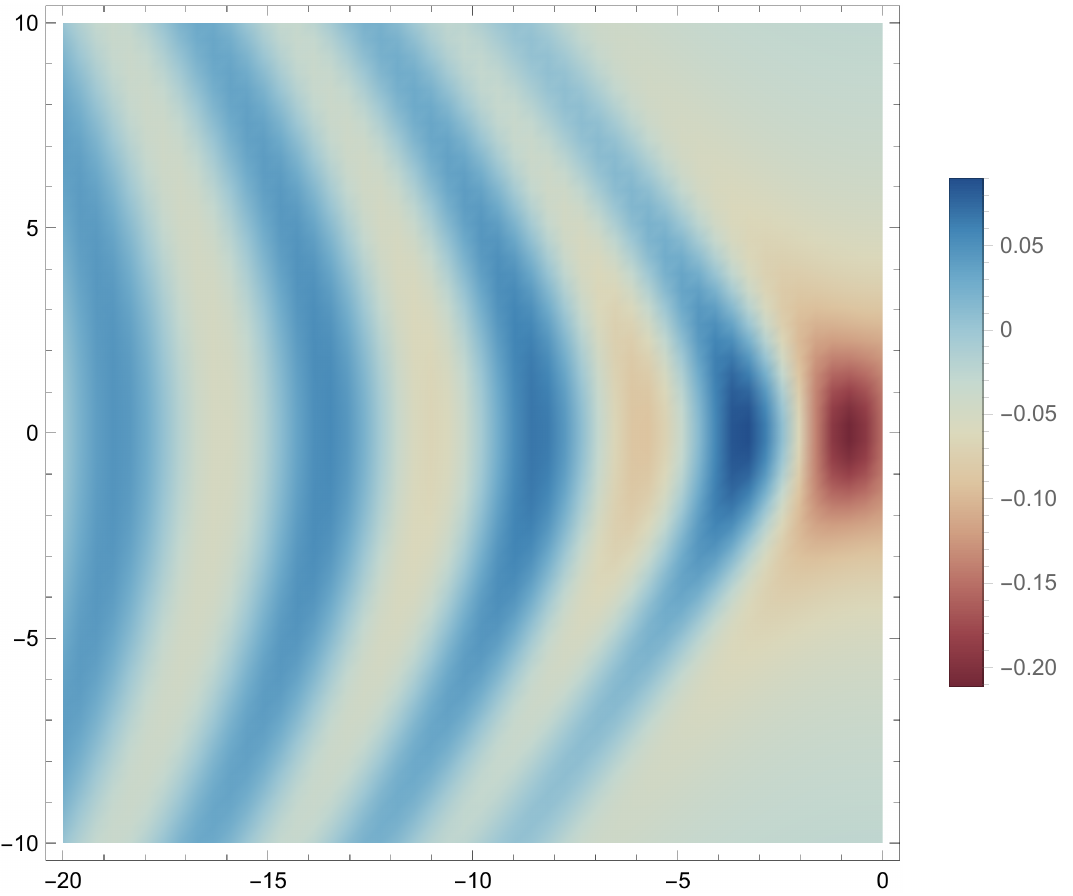} 
		\par\end{centering}
	\centering{}\centering{}\includegraphics[width=8cm]{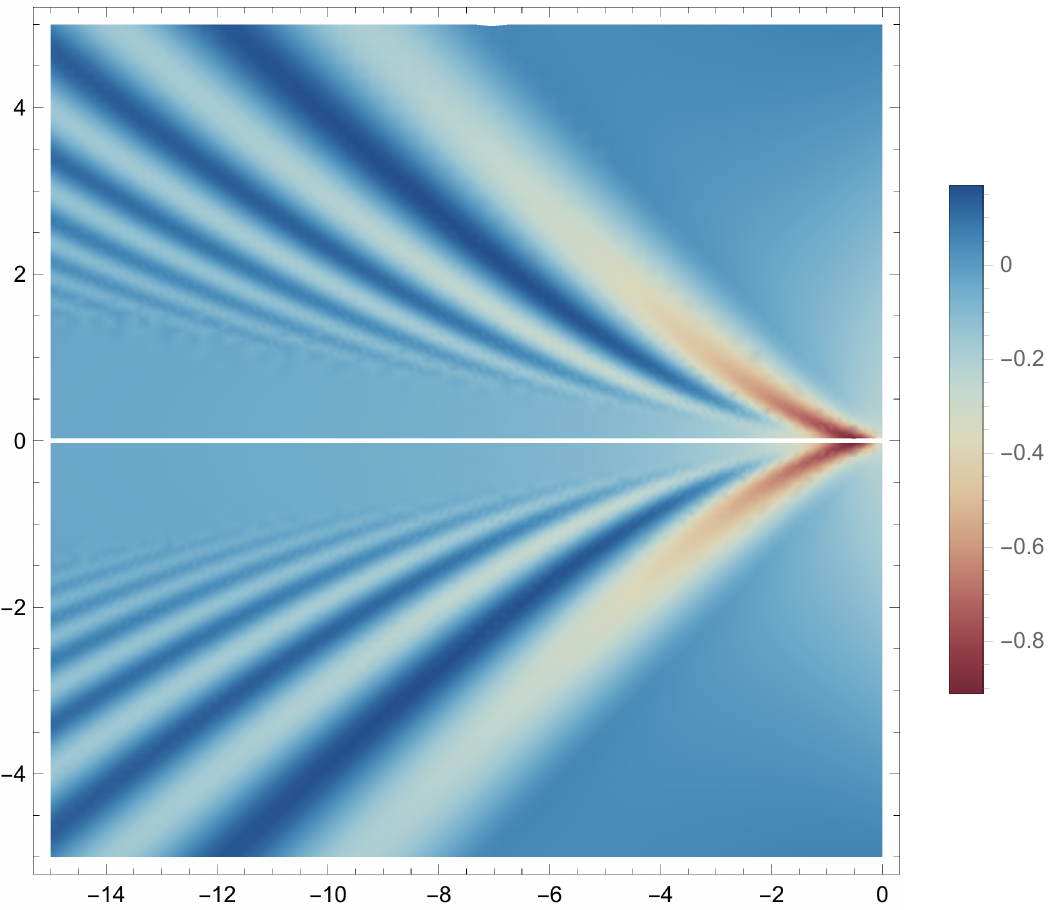}
	\caption{Electrostatic potential, in units of $\Phi_0$, in graphene for a particle moving parallel
		to it at $z_{0}=1\,{\rm \mu}$m with speed $v=0.1c$ (top panel; $E_{F}=0.17$\,eV) and computed using equations
		(\ref{eq:large_alpha}) and (\ref{eq:I1_small_Froude}),
		and at $z_{0}=0.1\,{\rm \mu}$m with speed $v=0.1c$ (bottom panel;
		$E_{F}=0.17$\,eV) and computed using equations (\ref{eq:small_alpha}) and (\ref{eq:I1_large_Froude}). Note the quantitative agreement between corresponding panels in this figure and figure \ref{fig:Kelvin}.  The axes of the figures are in $\mu$m.
		The electrostatic potential was computed using the formula $\Phi_{\rm in}/\Phi_0=I_1+\Re I_2/(4\pi^2)$.
		\label{fig:wake_asymp}}
\end{figure}

To compare the asymptotic expressions to the exact results
we depict  in figure \ref{fig:wake_asymp} the same wakes shown in figure \ref{fig:Kelvin} but computed using equations 
(\ref{eq:large_alpha}) and (\ref{eq:I1_small_Froude}), and
equations
 (\ref{eq:small_alpha}) and (\ref{eq:I1_large_Froude}).
It is clear that the shape of the approximated wakes given in figure \ref{fig:wake_asymp} is in qualitative agreement with the wakes depicted in figure \ref{fig:Kelvin}. The agreement between the wake for $z_0=1\,\mu$m is only qualitatively accurate since the
ratio $az_0/v^2\simeq 1.2$ is not in the regime $az_0/v^2\gg 1$.
Had we chosen a larger $z_0$ or a smaller $v$ and the agreement between the two wakes would have also been quantitatively better. In particular we note the disagreement in the value of the opening angle of the wake's cone (larger in the approximate wake).
In the case of the bottom panel of figure \ref{fig:wake_asymp} the agreement with the bottom panel of figure \ref{fig:Kelvin}
is quite good, since in this case  the wake for $z_0=0.1\,\mu$m is more accurate, given that the
ratio $az_0/v^2\simeq 0.12$ can be considered to fulfill the condition $az_0/v^2\ll 1$. Indeed, in both cases the cone of the wake  has the same coordinate $y\approx \pm2\,\mu$m, for 
$x=-15\,\mu$m. It is also clear that the form of the wake in this
regime differs considerably from the previous one. This is a 
striking manifestation of the two aforementioned regimes. 
The existence of these two regimes was put in evidence in the 
numerical studies of table \ref{tab:Estimation-of-half-angle} 
and figure \ref{fig:transition}.

It is clear from this analytical analysis that there is a transition
in the shape of the wake around $az_0/v^2\sim 1$ which is 
precisely the parameter that enters in the formula (\ref{eq:theta_guess}), derived based on intuition, numerical,
and dimensional analysis. Therefore the regime $az_0/v^2\gg1$ defines the Kelvin-like behavior of the wake, whereas the opposite regime defines the Mach-like behavior.

If we zoom out the wake in the top panels of Figures 
\ref{fig:Kelvin} and \ref{fig:wake_asymp}, we can clearly identify the presence of a plane wave superimposed on the wake. 
This is also evident in the bottom panel of figure \ref{fig:Kelvin}.
This plane 
wave presents a number of crests and valleys. 
We note  that the number of nodes and crests in the 
approximated wakes  coincide with the same quantity in the exact wakes. Indeed if  in equation
(\ref{eq:large_approx}) we make the approximation
(since in this case the integral is dominated by values of $u\approx0$)
\begin{equation}
	\sin\left(
	\frac{ar}{v^2}\cos\theta^\prime(1+u^2/2)
	\right)\approx \sin\left(
	\frac{ar}{v^2}\cos\theta^\prime
	\right)
\end{equation}
 it is then clear that we have superimposed to the wake pattern a plane wave of the form $\sin(ar\cos\theta^\prime/v^2)=\sin(ax/v^2)$. Considering the 
case of the top panel of figure \ref{fig:wake_asymp}
we have the ratio $a/v^2\approx1.2\,\mu$m$^{-1}$ . Therefore the wavelength of the wave reads $\lambda\approx 2\pi v^2/a\approx 5\,\mu$m, meaning that in the distance $\Delta x=20\,\mu$m we should have four crests, which is exactly what is seen in the
top panels of Figures \ref{fig:Kelvin}  and \ref{fig:wake_asymp}.
The  number of crests and valleys seen in the 
bottom panel of figure \ref{fig:Kelvin} is three, since 
$\Delta x=15\,\mu$m .
We note the absence of the plane wave pattern in the 
wake of the bottom panel of figure \ref{fig:wake_asymp}; 
this is a consequence of the approximation of the argument 
of the sine-function used in the limit $az_0/v^2\ll 1$ (we have used a large $u$ expansion and the plane wave depends on the small $u$ values, as seen in the previous equation).

\begin{figure}[t!]
	\begin{centering}
		\includegraphics[width=8cm]{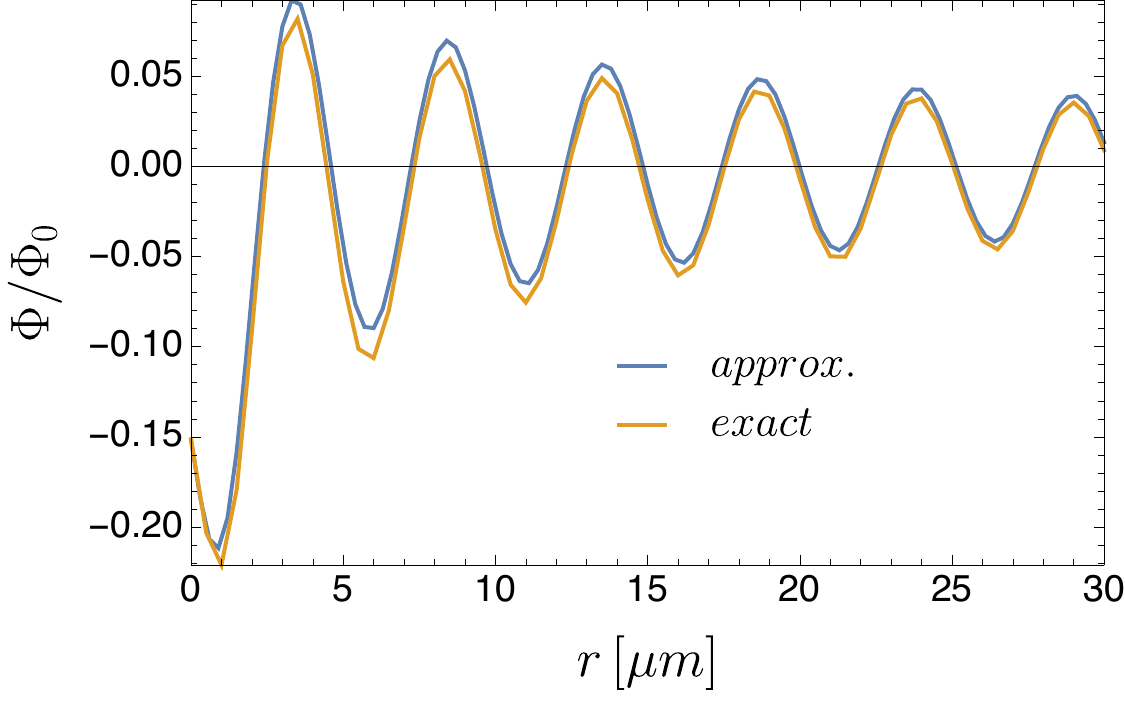} 
		\par\end{centering}
	\centering{}\centering{}\includegraphics[width=8cm]{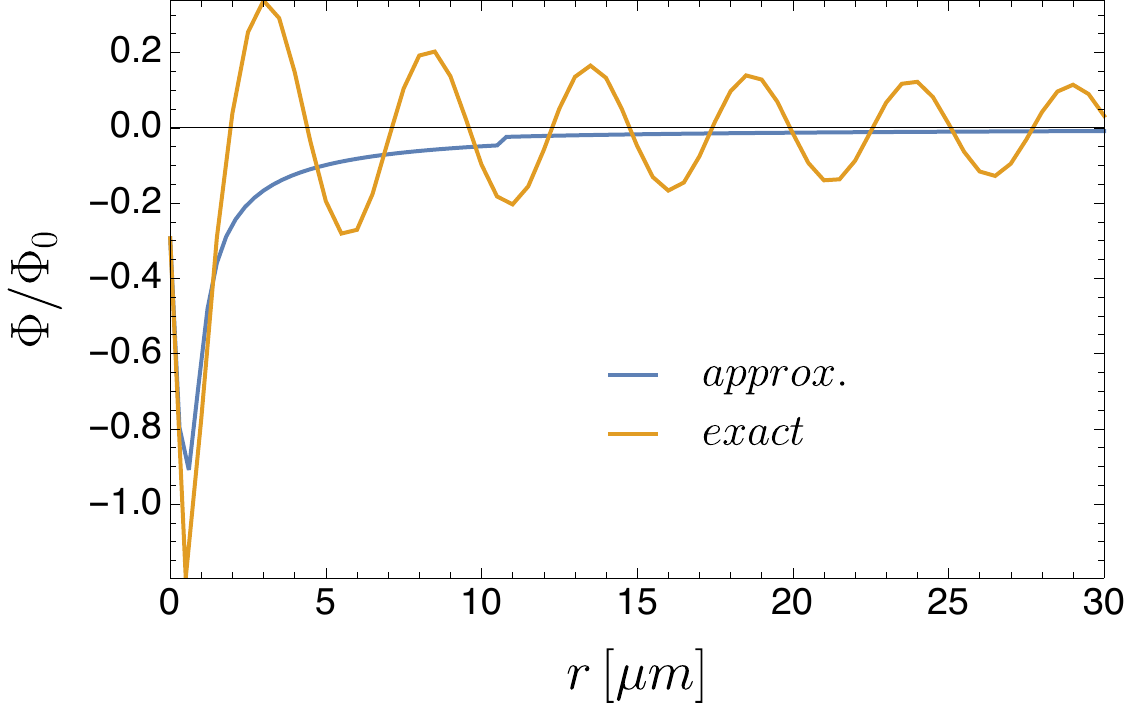}
		\centering{}\centering{}\includegraphics[width=8cm]{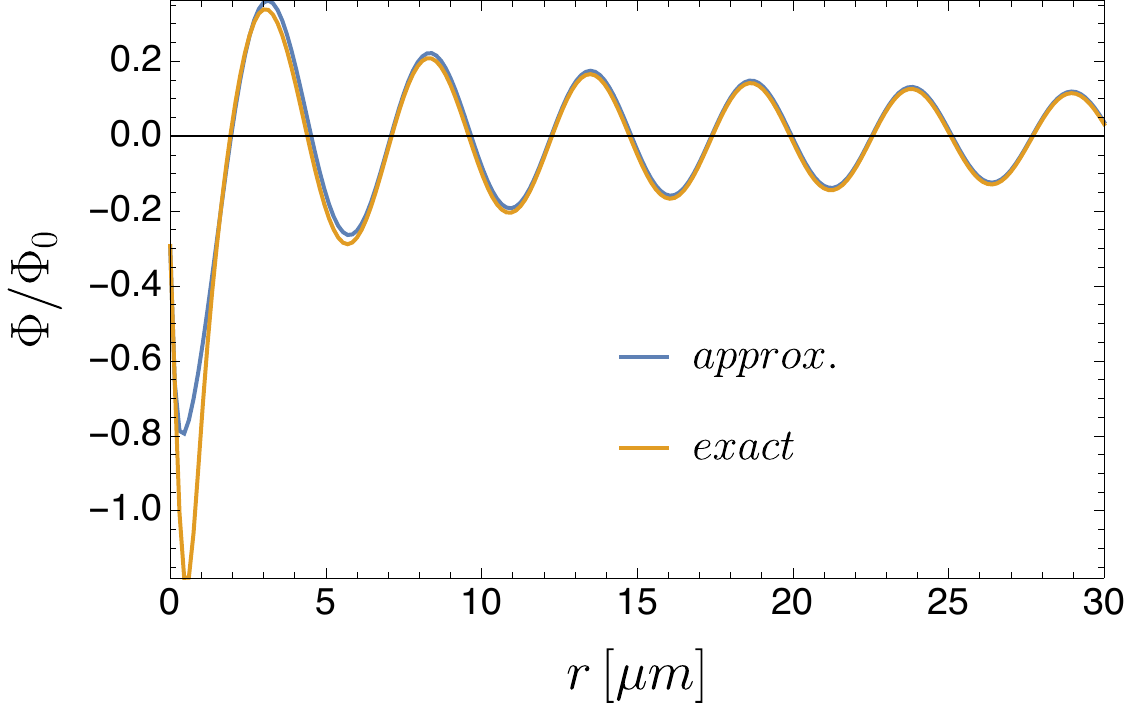}
	\caption{Comparison of the exact and approximated expression for the  potential, in units of $\Phi_0$, along the direction $\theta^\prime=\pi$.
		The parameters are the same as in figure \ref{fig:wake_asymp}. In the top panel de Froude number is $\rm {Fr_{pl}}=0.90$ and in the central and bottom ones is
		$\rm {Fr_{pl}}=2.86$. 
		Note the excellent agreement of the approximated result in the top panel, even though the Froude number is only slightly smaller than 1. 
		Also note that in the central panel, for moderate  Froude number, the plane wave is missing in the approximated curve [computed using 
		equations (\ref{eq:small_alpha}) and (\ref{eq:I1_large_Froude})]; the  reason for this is discussed in the text. In the bottom panel we improved over the 
		approximation of the central panel using equations (\ref{eq:large_alpha}) and (\ref{eq:wake_intermediate}).
		However this level of approximation is only valid in the 
		regime $az_0/v^2\lesssim1$ where the plane wave is well developed. All the approximated curves were computed using the formula $\Phi_{\rm in}/\Phi_0=I_1+\Re I_2/(4\pi^2)$.
		  \label{fig:comparison_asymp}}
\end{figure}

In figure \ref{fig:comparison_asymp} we show the electrostatic potential, $\Phi/\Phi_0$, along the direction $\theta^\prime=\pi$ ($y=0$). The agreement between the exact and the approximated formulas is excellent. Note, in the central panel,  the missing plane wave oscillations in the approximated result for large Froude number; the reason for this  has been discussed already.  However, if we decrease $z_0$ for increasing the Froude number, the wavelength of the plane wave becomes very large and agreement between the approximated and numerically exact solutions is excellent.
The agreement between the approximated curves
and the numerical exact ones is also good for values of
$y\ne0$ (results not shown). The fact that the plane wave is missing in the limit of large Froude numbers hints that the approximation of  the 
arguments of the trigonometric functions by their values in the limit $u\rightarrow\infty$ is too drastic. This conclusion suggests keeping for the value of
$\Re I_2$ the same dependence has that used in the limit of small Froude number, since this part contains the plane wave, that is, we use equation (\ref{eq:large_alpha}) for $\Re I_2$. As for the integral $I_1$ we expand the
arguments of the two cosines differently: the one responsible for the plane wave is expanded in the limit of small $u$ whereas the other is expanded in the limit of large
$u$. With this procedure, we obtain for $I_1$ the result
\begin{align}
	I_1&\approx\frac{\sqrt{\pi/2}}{4\pi^2}
	\left[
	\frac{e^{-s_1/2}}{\sqrt{2\beta-i(\gamma-2\lambda)}}
	f(s_1/2)+{\rm c.\,c.}
	\right]\nonumber\\
	&+\frac{\sqrt{\pi/2}}{4\pi^2}
	\left[
	\frac{e^{-s_2/2}}{\sqrt{2\beta-i(\gamma+2\lambda)}}
	f(s_2/2)+{\rm c.\,c.}
	\right]\,,
	\label{eq:wake_intermediate}
\end{align}
with
$s_1=2 \beta -2 i \gamma +i \lambda$, $s_2=2 \beta -i (2 \gamma +\lambda )$, and
\begin{equation}
f(z)=i\pi{\rm erf}\left[i\sqrt{\vert z\vert}e^{i{\rm arg}(z)/2}\right]\,,
\end{equation}
where ${\rm arg}(z)$ is the argument of the complex number $z$ and $\Re z>0$.
This last result for $I_1$ together with equation (\ref{eq:large_alpha}) 
prove to be accurate in the regime $az_0/v^2\lesssim1$, as seen in the bottom panel of figure \ref{fig:comparison_asymp}.  Therefore this approach includes the plane wave present in the wake at intermediated Froude numbers. However, being exceptionally good in describing the $\theta^\prime=\pi$ case, this approximation does not
excel for moderate to large angular deviations from $\theta^\prime=\pi$, as it underestimates the amplitude of the plane wave along these directions.

From the previous  analytical study,
we learn that the plane wave existing in the wake has its wavelength controlled by the value of $v^2/a$ and $z_0$
plays no immediate role in this.
 From the analytical solution we also learn that the effect of the moving charge on the plasmonic wake diminishes exponentially with $z_0$ due to the exponential factor
$e^{-az_0/v^2}$. Therefore the regime $az_0/v^2\gg1$ is likely 
to be experimentally challenging as the charge fluctuations are exponentially suppressed. This is also clear from the vertical scales of figure \ref{fig:wake_asymp}. From the previous discussion, it is obvious that the dimensionless ratio 
$z_0a/v^2$ plays a fundamental role in determining the nature of the wake. As argued in section \ref{sec:Conclusions} this quantity is related to the Froude number of plasmonic wakes induced by the Coulomb dragging effect of the passing charge.

Equation (\ref{eq:small_alpha}) can also be used to motivate equation (\ref{eq:theta_guess}). The procedure is somewhat delicate and we only outline the 
main steps. Firstly, we expand this equation in powers of the Froude number followed by an expansion in powers of $1/\sqrt{r}$ when $r\rightarrow\infty$ 
and gather the terms that decay slower with $r$ ---those proportional to $1/\sqrt{r}$ (the other terms are discarded, that is, we make the far field approximation). Secondly, we expand the resulting function (the amplitude of the spatial dependent trigonometric function) around $\theta^\prime=\pi$ and obtain a function $f(\theta^\prime)$. Since we seek the maximum of the potential, we take the derivative of  $f(\theta^\prime)$ 
and equal it to zero, $f^\prime(\theta^\prime)=0$. Solving the previous equation for $\theta^\prime$ gives the position of the first maximum of the potential, which within the approximation of  equation (\ref{eq:small_alpha}), reads $\theta^\prime=\pi\pm \sqrt{5}/{(2\rm Fr_{pl}})$, which has the correct order of magnitude we found from the fit made in figure \ref{fig:transition}.

\section{ A source of  graphene plasmons\label{sec:nano-rectangle}}

\begin{figure}
	\centering{}\centering{}\includegraphics[width=8cm]{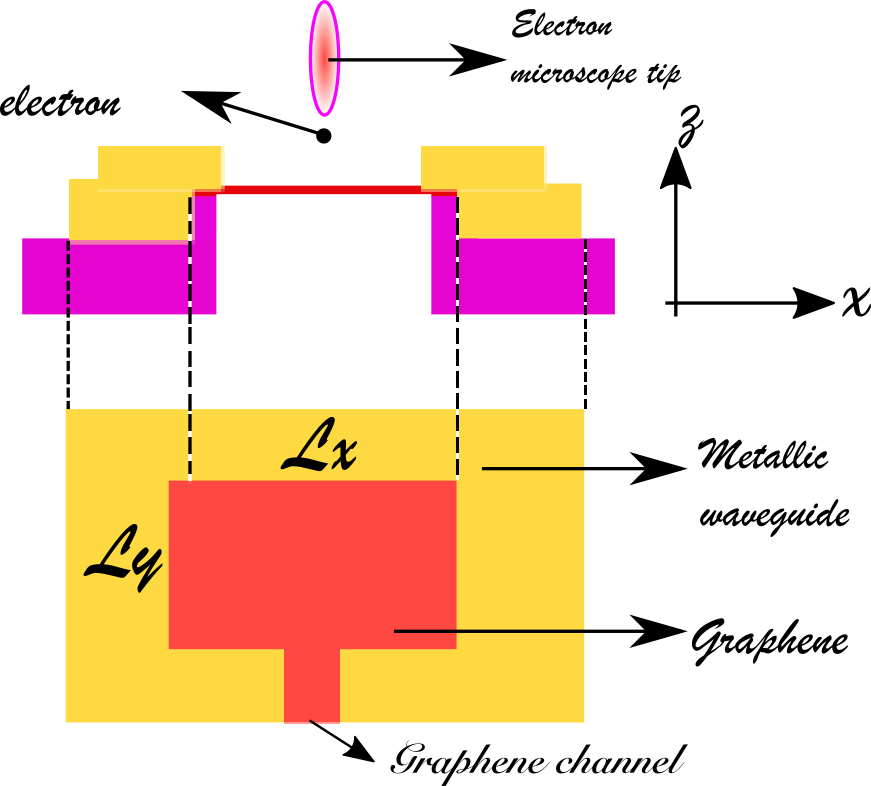}\caption{\label{fig:waveguide}
	A graphene sheet  enclosed in a metallic waveguide and suspended 	over a trench. The tip of an electronic  microscope injects electrons in the waveguide and the passing through electron excites localized plasmons in graphene. 
	If one bores a hole ---graphene channel--- in the metallic waveguide and let graphene occupy the hole and connects it to an external graphene sheet, this architecture can act as a source of graphene plasmons.}
\end{figure}

Next, we go back to the problem of a charged particle moving perpendicularly to a graphene sheet.
In this section we consider a rectangular waveguide, of cross-section area $A=L_xL_y$, where $L_x$ and $L_y$ are the sides of the rectangle, as depicted in figure~\ref{fig:waveguide}.
A micro-rectangle of graphene is in the middle of the waveguide and an electron is sent along the axis of the waveguide. The waveguide will support discrete graphene plasmonic modes, which can be excited by the passing electron. The goal of this section is to determine the EEL spectrum of the micro-rectangle of graphene. This method will allow to probe and excite discrete plasmonic resonances in graphene. 

This architecture can also be used as a source of plasmons. The idea is conceptually simple: a hole is bored in the metallic waveguide, such that a graphene ribbon extends itself outside the waveguide and connects to an external graphene sheet.
The plasmons, once excited in the suspended graphene \emph{drum skin} will propagate away from the drum though the channel connecting the drum to the external graphene sheet. Choosing drums of different shapes and sizes allows to span a vast spectral range of graphene plasmons.  Note that the graphene drum supports in-plane oscillations (compressible charge-density waves), rather than out-of-plane displacements (of the skin) common to the classical sound drum.

For solving this problem we have to consider, in addition to the hydrodynamic model, equations (\ref{eq:hydrodinamic_model}), the boundary conditions introduced by the waveguide walls, that is:
\begin{subequations}
\begin{equation}
\left.\frac{\partial \Phi}{\partial x} \right\vert_{y=0}=\left.\frac{\partial \Phi}{\partial x} \right\vert_{y=L_y}=0,
\end{equation}
\begin{equation}
\left. \frac{\partial \Phi}{\partial y} \right\vert_{x=0}=\left. \frac{\partial \Phi}{\partial y} \right\vert_{x=L_x}=0,
\end{equation}
\begin{equation}
v_y(y=0)=v_y(y=L_y)=v_x(x=0)=v_x(x=L_x)=0, \label{eq:velocity_waveguide}
\end{equation} \label{eq:boundary_conditions}
\end{subequations}
that is, we consider a perfect metallic conductor so that the tangential component of the electric field is null at the waveguide walls and the perpendicular component of the electronic current is null at the graphene boundary. 

\subsection{Discrete plasmon dispersion}

Firstly we consider the plasmonic solutions of the hydrodynamic model. The potential can be conveniently expanded in a Fourier series as:
\begin{equation}
\Phi_{\rm{in}}(x,y,z)=\sum_{n,m=1}^\infty A_{nm}\sin\left(\frac{n\pi x}{L_x} \right) \sin \left( \frac{m \pi y}{L_y}\right)e^{-k_{nm} |z|}, \label{eq:induced_field}
\end{equation}
where
\begin{equation}
k_{nm}=\sqrt{ \left(\frac{n\pi}{L_x}\right)^2+\left( \frac{m \pi}{L_y} \right)^2},
\end{equation}
and integrating Poisson's equation  (\ref{eq:poisso_2d_density}) with respect to the $z$ coordinate we have:
\begin{align}
n_1(x,y,\omega)&=-\frac{2 \varepsilon_0}{e}\sum_{n,m=1}^\infty  A_{nm}(\omega) k_{nm}
\nonumber\\ 
&\times \sin\left(\frac{n\pi x}{L_x} \right) \sin \left( \frac{m \pi y}{L_y}\right),  \label{eq:density_drum}
\end{align}
and the velocity components can be calculated with equations (\ref{eq:continuity_linear}) and (\ref{eq:velocity_waveguide}):
\begin{subequations}
\begin{equation}
v_x(x,y)=\frac{2 \varepsilon_0}{e}\frac{i\omega}{n_0} \sum_{n,m=1}^\infty B_{nm}
\cos\left( \frac{n \pi x}{L_x} \right) \sin  \left( \frac{m \pi y}{L_y} \right),
\end{equation}
\begin{equation}
v_y(x,y)= \frac{2 \varepsilon_0}{e}\frac{i\omega}{n_0} \sum_{n,m=1}^\infty C_{nm} \sin\left( \frac{n \pi x}{L_x} \right) \cos  \left( \frac{m \pi y}{L_y} \right),
\end{equation} \label{eq:velocity_expression}
\end{subequations}
with the amplitudes $A_{nm}$, $B_{nm} $, and $C_{nm}$ related via
\begin{equation}
A_{nm} k_{nm}=B_{nm} \frac{n\pi}{L_x}+C_{nm}\frac{m\pi}{L_y}.
\end{equation}

Lastly using equation (\ref{eq:hydrodinamic_model}) it follows the discretized version of the
plasmon dispersion relation (\ref{eq:omega_spp}):
\begin{equation}
2\alpha E_F \hbar c k_{nm}+\frac{v_F^2 \hbar^2}{2} k_{nm}^2=\hbar^2\omega_{nm}^2. \label{eq:discret_dispersion}
\end{equation}
Quite intuitively, this solution for the plasmons spectrum describes localized surface plasmons in the graphene sheet, in the form of standing waves.

\subsection{Motion of a charge along the axis of symmetry of the waveguide: Charge density fluctuation and  EEL spectrum}

Following the steps of section \ref{sec:perp_motion}, we consider the presence of an electron, moving parallel to the axis of the waveguide, and impinging perpendicularly on the graphene sheet. The density of external charge is given by:
\begin{equation}
\rho_{\rm{ext}}=Ze \delta(x-x_0) \delta(y-y_0) \delta(z-vt),
\end{equation}
that corresponds to a charged particle with velocity $v$ aimed to the point $(x_0,y_0)$ on the graphene surface.

The electric potential follows from solving Poisson's equation $\nabla^2 \phi_{\rm{ext}}=-\rho_{\rm{ext}}/\varepsilon_0$ with the boundary condition (\ref{eq:boundary_conditions}). It then follows:
\begin{align}
\Phi_{\rm{ext}}(x,y,z,\omega)&= \frac{4 Ze }{\varepsilon_0 L_xL_y v}\sum_{n,m=1}^\infty \phi_{nm}(z) \sin\left( \frac{n \pi x}{L_x}\right) 
\nonumber\\
&\times\sin \left( \frac{m \pi y}{L_y} \right), \label{eq:external_field_drum}
\end{align}
with
\begin{equation}
\left(\partial_z^2-k^2_{nm}\right)\phi_{nm}(z)=-u_{nm} e^{-i \omega z/v},   \label{eq:phi_nm}
\end{equation}
where we have used the completeness relation to rewrite the Dirac delta function as
\begin{equation}
\delta(x-x_0)=\frac{2}{L_x} \sum_{n=1}^\infty \sin\left(\frac{n\pi x}{L_x} \right) \sin\left(\frac{n\pi x_0}{L_x} \right),
\end{equation}
 with a similar expression for $\delta(y-y_0)$, and we have  defined $u_{nm}=\sin(n\pi x_0/L_x) \sin(m\pi y_0/L_y) $.

Equation (\ref{eq:phi_nm}) can be solved with the Green's function method as in equation (\ref{eq:Greens_function}), leading to:
\begin{equation}
\phi_{nm}(z)=-\frac{ u_{nm} }{(  \omega/v)^2+k_{nm}^2}e^{-i \omega   z /v}.
\end{equation}

Taking the divergence of  Eq. (\ref{eq:external_hydrodynamic}), with the density given by the Fourier expansion (\ref{eq:density_drum}) and using the continuity equation after a Fourier transform in time we arrive
at:
\begin{align}
\frac{2\varepsilon_0}{e}\sum_{n,m=1}A_{nm}&\left(\omega^2-\omega^2_{nm} \right)k_{nm} \sin\left( \frac{n\pi x}{L_x} \right) \sin\left( \frac{n\pi y}{L_y} \right)=\nonumber \\
&-\frac{n_0 e v_F^2}{E_F} \nabla^2 \Phi_{\rm{ext}}(z=0),
\label{eq:Anm}
\end{align}
with $\omega_{nm}$ given by Eq.~(\ref{eq:discret_dispersion}). Finally, after projecting equation (\ref{eq:Anm}) in one of the basis functions, we find:
\begin{equation}
A_{nm}= 8\Phi_{0}\frac{v}{L_xL_y}\frac{ v_F^2}{E_F}\frac{k_{nm}}{\omega^2-\omega^2_{nm}}\phi_{nm}(0).
\end{equation}

%\begin{equation}
%A_{nm}= \frac{2 e^2 Z}{\varepsilon_0^2 a b v}\frac{n_0 e v_F^2}{E_F}\frac{k_{nm}}{\omega^2-\omega^2_{nm}}\phi_{nm}(0).
%\end{equation}

The knowledge of $A_{nm}$ allows to first compute $n_1(x,y,\omega)$ and from this the determination of $n_1(x,y,t)$ is possible after a Fourier transform.
The determination of the velocity field requires the knowledge of the coefficients $B_{nm}$ and $C_{nm}$ which can be calculated using the density (\ref{eq:density_drum}), the velocity (\ref{eq:velocity_expression}),  the electrostatic potential (\ref{eq:induced_field}) and (\ref{eq:external_field_drum}), into
the equation (\ref{eq:wave_equation_vector}) and using the dispersion relation (\ref{eq:discret_dispersion}), giving:
\begin{subequations}
	\begin{equation}
		B_{nm}= \frac{n\pi}{L_x}\frac{A_{nm}}{k_{nm}},
	\end{equation}
	\begin{equation}
		C_{nm}=\frac{m\pi}{L_y}\frac{A_{nm}}{k_{nm}}.
	\end{equation}
\end{subequations}
Note that equation (\ref{eq:wave_equation_vector}) is vectorial, thus allowing for the determination of the two previous coefficients.
%
%Finally the induced charge read:
%\begin{equation}
%n_1(x,y)=\Phi_0\sum_{nm=1}^\infty \frac{k^2_{nm}\phi_{nm}(0)}{\omega^2-\omega_{nm}^2}  \sin\left(\frac{n\pi x}{a}\right) \sin\left(\frac{m\pi y}{b} \right),
%\end{equation}
\begin{figure*}
	\centering{}\centering{}\includegraphics[width=16cm]{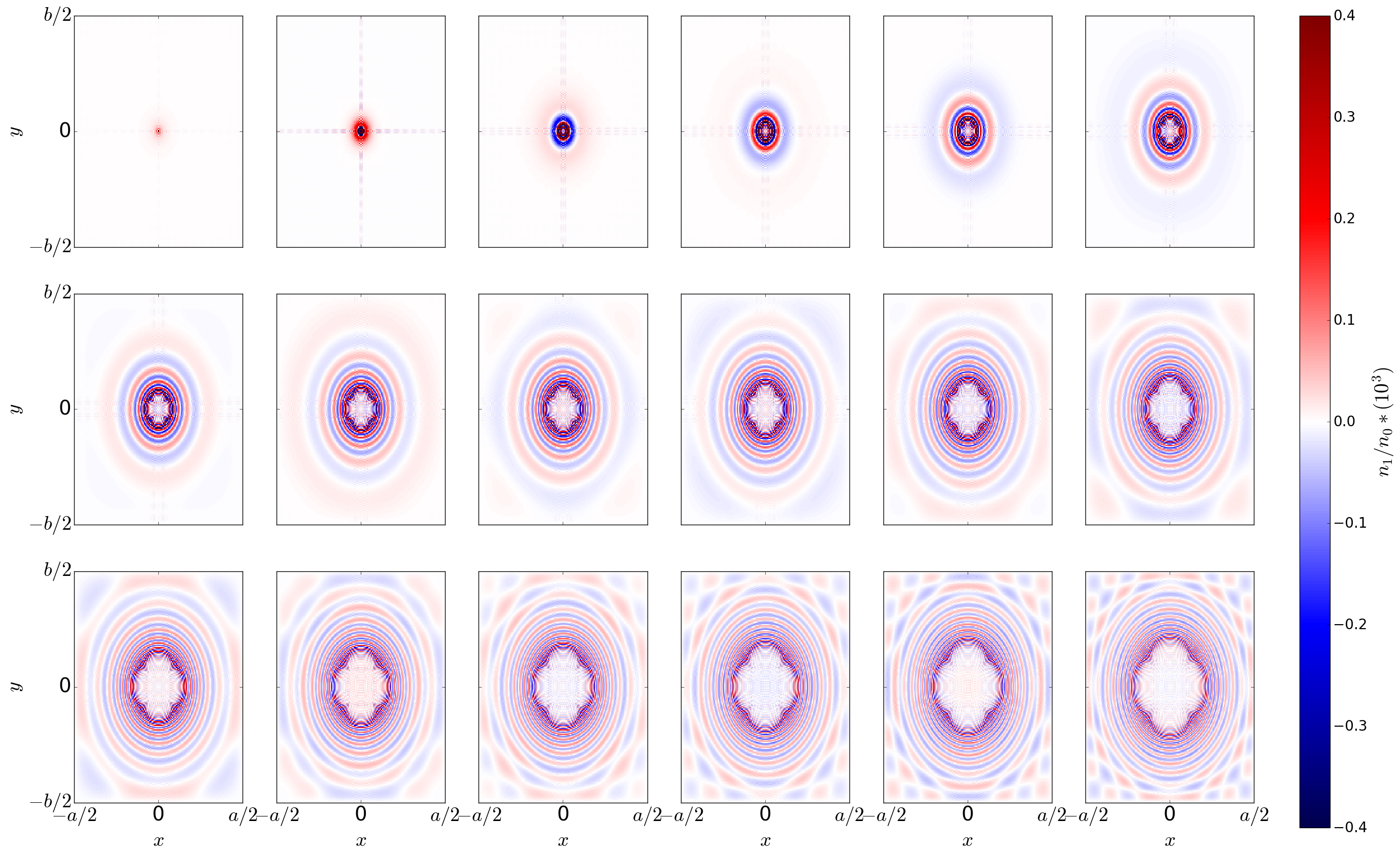}\caption{\label{fig:Fig_n1_a2b2}
		Induced electronic density fluctuations $n_1(x,y)$ in the graphene drum for $t=0.1$\,fs$+18(n-1)$\,fs, where $n$ is the number of the panel (ordered by rows from left to right). The parameters are: $v=0.1c$, $E_F=0.37$\,eV, and $L_x=L_y=3\,{\rm\mu}$m.  Note the evolution of the density from a circle for short times and the development of interference fringes for larger times. The blue and red colors correspond, respectively, to negative ad positive charge density fluctuations $n_1(x,y)$ relatively to the homogeneous charge density $n_0$. Note that the wavelength of the plasmons when they are first created is much smaller than the size of the drum. It is therefore conceivable that they can escape through a graphene channel drilled in the walls of the waveguide. The figures present a high degree of symmetry because the impact point chosen for the electron is the center of the square.}
\end{figure*}

In figure~\ref{fig:Fig_n1_a2b2} we depict the electronic density $n_1(x,y,t)$ on the drum, for ascending times, induced by an electron transversing graphene at a speed of $v=0.1c$ (kinetic energy of 2.5\,keV; note that this energy value is of the order of the energy used in SEM imaging of graphene and therefore would cause little to none damage to the material. We stress that the knock-on-threshold
for electrons in graphene is much larger, close to 80\,keV). Note in this figure the evolution of the charge density from initially circular concentric waves at the center of the drum for shorter times to interference fringes at larger times, due to reflection at the boundaries of the drum.  In the presence of the channel (see figure \ref{fig:waveguide}), the plasmonic wave reaches the boundary of the drum and some of the plasma frequency components will propagate through the channel outwards. Note that in figure \ref{fig:Fig_n1_a2b2} the electron hits the drum at its center. By choosing a different impact point we may relax reflection symmetries of the problem, thus in turn producing more directional waves (results not shown). In passing, we note that such directional waves could also be explored in non-integrable geometries, such as chaotically shaped billiards, or by turning from graphene to anisotropic 2D materials.

The EEL spectrum can be calculated using the same definition and method used in the previous section, and reads:
\begin{equation}
\Gamma(\omega)=\frac{\Psi_0}{\omega}\sum_{n,m=1}^\infty \delta(\omega-\omega_{nm}) \frac{k_{nm}^2 u_{nm}^2}{ [\omega^2+(k_{nm} v)^2]^2},
\label{eq:EELS_drum}
\end{equation}
with
\begin{equation}
\Psi_0=  32 \pi^2 \alpha^2 \hbar c^2  \frac{n_0 v_F^2 v^2}{L_xL_y E_F}.
\end{equation}
The emergence of Dirac delta functions in the EEL spectrum guaranties that a single electron can excite multiple plasmon modes, albeit with different weights.

In the limit $L_x=L_y\rightarrow\infty$ the sums are conveniently converted into an integral:
\begin{equation}
\Gamma(\omega)=\frac{\Psi_0}{\omega}
\frac{L_xL_y}{\pi^2}\int_0^{\pi/2}d\theta\int_0^\infty kdk
\delta(\omega-\omega_{k})\frac{k^2u^2(k,\theta)}{[\omega^2+(k v)^2]^2}
\end{equation}
We have to perform the integral over the delta function, which is elementary, and the angular integral, leading to: 
\begin{align}
\Gamma(\omega)&=\Psi_0
\frac{L_xL_y}{\pi^2}\int_0^{\pi/2}d\theta \frac{2\omega^2u^2(\omega^2/a,\theta)}{(a^2+v^2\omega^2)^2}\nonumber\\
&=\frac{2\hbar}{ E_F}
\frac{v^2\omega^2/a^2}{\left(1+v^2\omega^2/a^2\right)^2}
\end{align}
where, as before, we have written $\omega_k=\sqrt{a k}$ (recall that $a$ has unit of acceleration) and used the limit (choosing $x_0=y_0=L_x/2$)
\begin{equation}
\lim\limits_{L_x\rightarrow\infty}\frac{16}{\pi}
\int_{0}^{\pi/2}d\theta u^2(\omega^2/a,\theta)=2\,.
\end{equation}
We have therefore recovered the result for a continuous graphene sheet, given by equation (\ref{eq:EELs_simple}).

\begin{figure}
	\centering{}\centering{}\includegraphics[width=8cm]{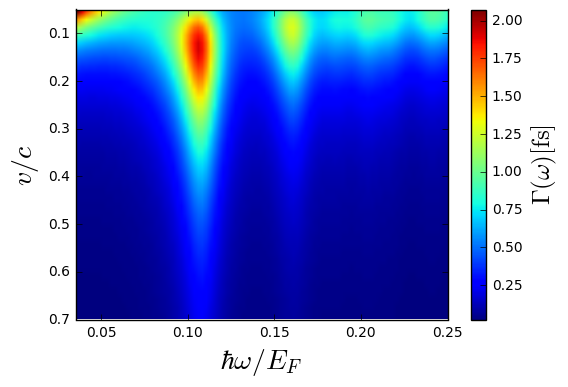}
	\caption{\label{fig:EELS_drum}
		EEL spectrum of a graphene square drum ($L_x=L_y=3\,{\rm\mu}$m) as function of the speed of the moving charge and of the frequency. If we draw a horizontal line throughout the figure we obtain the EEL spectrum for a given speed. It is clear that this cut produces a new figure with peaks at certain frequencies, corresponding to the excitation of the localized plasmons in the drum. We have considered a Fermi energy of $E_F=0.37$\,eV and have broaden the delta functions to Lorentzians with a relaxation rate of 4.1\,meV.}
\end{figure}

In figure \ref{fig:EELS_drum} we depict the EEL spectrum of a 
graphene drum, given by equation (\ref{eq:EELS_drum}). It is clear from this figure that a single electron can excite more than one surface plasmon mode. This is evident from the presence of several peaks in the EEL spectrum at different
frequencies. This effect has already been seen in the excitation of localized plasmons in graphene nano-structures \cite{key-18}. Also note that there is an optimal speed for the more efficient excitation of  plasmons. For the first peak (at low frequencies) the optimal speed is in the interval $v\in[0.1c,0.2c]$.

%-----------------------------------------------------------%
\section{Conclusions\label{sec:Conclusions}}

We have considered the problem of the excitation of surface plasmons
in graphene by a fast moving charge. We have analyzed two cases: (i)
when the charge is moving perpendicular to the graphene sheet and
(ii) when the charge is moving parallel to the graphene sheet. In
the first case we have computed the EEL spectrum~\cite{key-28} and
found that surface plasmons from a continuum of wave numbers are excited
by this method. The excitation of surface plasmons of different frequencies
has been discussed in the literature before~\cite{key-18} in the
context of graphene based nano-structures. In this reference it is
shown that for continuous graphene the EEL spectrum has a maximum
at (in agreement with our result) $\hbar\omega/E_{F}=2\alpha c/v$
implying that the frequency maximum shifts to higher frequencies as
$v$ decreases, as seen in figure \ref{fig:Loss-spectrum}. Taking,
for example, $v/c=0.01$, the previous result implies that $\hbar\omega/E_{F}\approx1.5$,
which agrees with figure~\ref{fig:Loss-spectrum}. As we have seen,
in the case of an infinite graphene sheet the EEL spectrum has broad
resonances, which is an indication of the simultaneous excitation
of plasmons of different energies. In the second case, the moving
charge induced a plasmonic  wake as it moves over graphene. 

Contrary
to what could have been expected, the wake induced by the moving charge is not necessarily of Kelvin type
and can also be of Mach type, where the angle of the cone of the charge
wake is proportional to the inverse of the speed of the moving charge
and therefore can be much smaller than the value predicted by Kelvin
theory. We have shown that there is a transition Froude number  from Kelvin
to Mach-type of ship wake. These two different possibilities have
been observed in real ship wakes and the transition  is controlled by
the Froude number~\cite{key-4}, ${\rm Fr}= \sqrt{U^2/(Lg)}$, where $U$ is the ship speed, $L$ is the hull ship length, and $g$ is the acceleration of gravity.  We have noted in our graphene problem that the 
dimensionless number $az_0/v^2$ determines the transition
from the Mach-like to Kelvin-like regimes. Therefore, we
interpret  ${\rm Fr_{pl}}=\sqrt{v^2/(z_0a)}$ as the Froude number for the plasmonic wakes in graphene, where the "acceleration" $a$
depends on the Fermi energy of graphene, and controls the dispersion of graphene plasmons. The value ${\rm Fr_{pl}}\sim2$
 defines the transition region from Mach-type,
${\rm Fr_{pl}}\gg1$, to Kelvin-type, ${\rm Fr_{pl}}\ll1$, regimes.
Indeed, for large Froude number in ship wakes the half angle of the aperture of the cone is given by~\cite{key-4} 
$\theta_{\rm ship}\approx(2\sqrt{2\pi}{\rm Fr})^{-1}$ which 
would predict a $\delta=2\sqrt{2\pi}$, a number that is  a
 much larger the value $\delta\approx1.09$ (for $\theta$ in radians) we have found 
from our simulations. 
Presumably, our equation (\ref{eq:theta_guess}) is the first term of a series, in powers of $1/{\rm Fr_{pl}}$, of a more
complex expression involving the Froude number (see reference ~\cite{key-4} for the case of ship wakes). We have also gathered numerical evidence of a critical Froude number that signals the transition from Mach-type to Kelvin-type of wakes, whose value 
reads approximately $\rm {Fr}_{pl}^c\approx2$. 
This value is 
about 4 times larger the critical Froude number in ship wakes~\cite{key-4}. On the other hand,
for point like objects immersed in a fluid, the critical Froude number
has been found to be larger than 2 \cite{apparent-angle}, in agreement with our results.
We note that the way the apparent opening angle of the wake is measured does not influence the value of the Froude number at which the transition occurs.
See reference \cite{apparent-angle} for a different  method from ours of measuring the angle.
Interestingly, the parameter $a$ also depends on the dielectric constant of the environment.
This introduces an additional degree of freedom, besides the tunning of the Fermi energy, to gain control over the Froude number, a situation that has no parallel in ship wakes.
We end this discussion noting that the group velocity of the plasmons in graphene reads $v_g=d \sqrt{ak}/dk=\frac{1}{2}\sqrt{a/k}=\frac{1}{2}\sqrt{a\lambda_{\rm sp}/(2\pi)}$. On the other hand
gravity waves in deep waters propagate with a group velocity $v_{w}=\frac{1}{2}\sqrt{g\lambda/(2\pi)}$. We note that $v_g$ and $v_w$ are identical, except that
in $v_g$ we have $\lambda_{\rm sp}$ the wavelength of the surface plasmon. It is therefore not so surprising that the same physics we find in ship wakes has also been
found in surface plasmon wakes as long as the distance of the charged particle to 
the graphene sheet $z_0$ plays the same role as the hull ship length $L$. This latter possibility was not evident from the outset.

Our calculations are valid for suspended graphene. It would
be interesting to extend them for graphene on hexagonal Boron Nitride
for discussing the excitation of phonon-plasmon-polaritons. Also,
extending this work to multilayer graphene \cite{key-Rad_transition}
is a natural continuation of this work. 

We have also discussed the formation of localized plasmons in a graphene drum in a metallic waveguide. As noted, the system can be used as a 
plasmon source of collimated beam. We explored the rectangular drum, but a circular drum is also feasible and it will 
originate plasmons of different frequencies. We note that since for waveguides of the order of $1\,\mu$m its cut-off frequency is in the near-IR, the electron,
once in the waveguide, can only  radiate above this cut-off frequency. Therefore, all radiation that appears below this frequency is plasmonic in nature and there will be no transition radiation in  that frequency range, at least for deep enough waveguides, compared with the plasmon confining length in the transverse direction. 
Also, we have studied the case where the electron impinges at the center of the square. If we had considered a different impinging point then we would have created highly directional plasmons oriented toward the channel that conducts the plasmons out of the drum.
If below graphene a metal is positioned 
at a distance of one or two layers of hexagonal Boron Nitride, then graphene will support strongly confined plasmons, akin to 
acoustic plasmons in a continuous graphene sheet. This setup 
would functioned as a source of acoustic plasmons.

We can consider the graphene drum a plasmonic billiard, which presents a characteristic spectrum distribution as function of frequency. Both in classical and quantum billiards, it is well known that depending on their geometry, the trajectories (classical) and spectrum (quantum) can be chaotic. It would be interesting to study the spectrum distribution of plasmonic billiards
in the future.
 
 \vspace{1cm}
 
{\it Note added:} after the submission of this paper we became aware of a similar work \cite{key-ship_wake_fluid}, but in the field of 
fluid dynamics. 

\begin{acknowledgments}
	A.J.C. acknowledges for a scholarship from the Brazilian agency CNPq
	(Conselho Nacional de Desenvolvimento Cientí­fico e Tecnol{\'o}gico).
N.M.R.P. acknowledges useful discussions with Jaime Santos and support from the European Commission through
the project ``Graphene-Driven Revolutions in ICT and Beyond\textquotedbl{}
(Ref. No. 696656) and the Portuguese Foundation for Science and Technology
(FCT) in the framework of the Strategic Financing UID/FIS/04650/2013.
The Center for Nanostructured Graphene (CNG) was financed by the Danish
National Research Council (DNRF103). N.A.M. is a VILLUM Investigator
supported by VILLUM Fonden. 
\end{acknowledgments}

%----------------------------------------------------------------------------------------------------------%
\appendix

\section{From Boltzmann equation to Euler's equation of hydrodynamics\label{sec:From-Boltzmann-equation}}

In this section we derive Euler's equation for fluid motion starting
from Boltzmann equation. Let us assume an electron gas characterized
by the distribution function $f(\mathbf{r},\mathbf{v})d\mathbf{r}d\mathbf{v}$,
which specifies the number of particles in the gas having position
and velocity centered at $\mathbf{r}$ and $\mathbf{v},$ respectively,
in the small volume $d\mathbf{r}$ and in the small velocity range
$d\mathbf{v}$. We can introduce a six-dimensional phase-space vector
$\mathbf{w}=(\mathbf{r},\mathbf{v})$ whose time rate reads $\dot{\mathbf{w}}=(\dot{\mathbf{r}},\dot{\mathbf{v}})$.
If the forces are conservative then $\dot{\mathbf{v}}=-\nabla\Phi$,
where $\Phi$ is the potential energy per unit mass. The time evolution
of the distribution function $f(\mathbf{r},\mathbf{v})$ is given
by the Boltzmann equation \cite{key-1} 
\begin{equation}
\frac{\partial f}{\partial t}+\mathbf{v}\cdot\frac{\partial f}{\partial\mathbf{r}}+\mathbf{g}\cdot\frac{\partial f}{\partial\mathbf{v}}=0
\end{equation}
where collisions have been excluded, and $\mathbf{g}$ represents
the external forces per unit mass. The previous equation is called
the collisionless Boltzmann equation. Adding collisions amounts to
adding a term of the form $\gamma(t)$ to the right-hand-side of this
equation. Boltzmann's equation is six-dimensional in phase space and
has more information that we actually need. Since we want to know
the position of the particles as function of time we can integrate
Boltzmann equation over the coordinate $\mathbf{v}$. Next we will
compute the first and second moments of the Boltzmann equation. To
that end we introduce the mass density using the relation~\cite{key-1}
(we are assuming all particles equal with mass $m$) 
\begin{equation}
\rho(\mathbf{r})=\int d\mathbf{v}f(\mathbf{r},\mathbf{v})m
\end{equation}
and the velocity moment via \cite{key-1} 
\begin{equation}
\langle v_{i}\rangle=\frac{1}{\rho(\mathbf{r})}\int d\mathbf{v}f(\mathbf{r},\mathbf{v})mv_{i}.
\end{equation}
It is also convenient to introduce the second moment of the velocity
as~\cite{key-1} 
\begin{equation}
\langle v_{i}v_{j}\rangle=\frac{1}{\rho(\mathbf{r})}\int d\mathbf{v}f(\mathbf{r},\mathbf{v})mv_{i}v_{j}.
\end{equation}
Let us now take the zero moment of the Boltzmann equation 
\begin{equation}
\int d\mathbf{v}\left[\frac{\partial f(\mathbf{r},\mathbf{v})}{\partial t}+\mathbf{v}\cdot\frac{\partial f(\mathbf{r},\mathbf{v})}{\partial\mathbf{r}}+\mathbf{g}\cdot\frac{\partial f(\mathbf{r},\mathbf{v})}{\partial\mathbf{v}}\right]=0
\end{equation}
from where it follows the continuity equation 
\begin{equation}
\frac{\partial\rho(\mathbf{r})}{\partial t}+\frac{\partial}{\partial\mathbf{r}}\cdot[\rho(\mathbf{r})\langle\mathbf{v}\rangle]=0
\end{equation}
stating mass conservation, and where we have used the divergence theorem
leading to the following identity 
\begin{equation}
\int_{V}d\mathbf{v}\frac{\partial f(\mathbf{r},\mathbf{v})}{\partial\mathbf{v}}=\int_{S_{\infty}}f(\mathbf{r},\mathbf{v})d\mathbf{S}_{\mathbf{v}}=0
\end{equation}
where $f(\mathbf{r},\mathbf{v})=0$ over a surface at infinity, $S_{\infty}$.

Let us next consider the first moment of the Boltzmann equation 
\begin{equation}
\int d\mathbf{v}\mathbf{v}\left[\frac{\partial f(\mathbf{r},\mathbf{v})}{\partial t}+\mathbf{v}\cdot\frac{\partial f(\mathbf{r},\mathbf{v})}{\partial\mathbf{r}}+\mathbf{g}\cdot\frac{\partial f(\mathbf{r},\mathbf{v})}{\partial\mathbf{v}}\right]=0
\end{equation}
which can be simplified to 
\begin{equation}
\frac{\partial}{\partial t}\left[\rho(\mathbf{r})\langle\mathbf{v}\rangle\right]+\sum_{i}\frac{\partial}{\partial x_{i}}\left[\rho(\mathbf{r})\mathbf{\langle v}v_{i}\rangle\right]-\mathbf{g}\rho(\mathbf{r})=0
\end{equation}
where the third term was computed using integration by parts. This
last equation is called the momentum equation. We now introduce the
tensor $\tau_{ij}^{2}=\langle v_{i}v_{j}\rangle-\langle v_{i}\rangle\langle v_{j}\rangle$.
This conveniently allows us to write the term with the second moment
in terms of products of first moments. Subtracting from the momentum
equation the continuity equation we obtain for each component 
\begin{equation}
\rho(\mathbf{r})\frac{\partial\langle v_{j}\rangle}{\partial t}+\rho(\mathbf{r})\sum_{i}\langle v_{i}\rangle\frac{\partial\langle v_{j}\rangle}{\partial x_{i}}=g_{j}\rho(\mathbf{r})-\sum_{i}\frac{\partial[\rho(\mathbf{r})\tau_{ij}^{2}]}{\partial x_{i}}
\end{equation}
or in vectorial terms 
\begin{equation}
\rho(\mathbf{r})\frac{\partial\langle\mathbf{v}\rangle}{\partial t}+\rho(\mathbf{r})(\langle\mathbf{v\rangle}\cdot\nabla)\mathbf{\langle v\rangle}=\mathbf{g}\rho(\mathbf{r})-\nabla P\label{eq:Euler}
\end{equation}
where $P$ is the pressure in the gas. The last equation is Euler's
equation of fluids dynamics and is the starting point for the hydrodynamic
model of plasmons in metals and in graphene. We should stress that
the derivation of equation~ (\ref{eq:Euler}) assumed a finite effective
mass $m$ for the particles in the gas. Adapting this equation for
graphene will require the introduction of graphene electrons' Drude
mass $m_{g}=\hbar k_{F}/v_{F}$, which links the Fermi momentum $\hbar k_{F}$
to the Fermi velocity $v_{F}$; both well-defined properties associated
with the linear dispersion of massless Dirac fermions in graphene.
This choice makes sense since we are describing transport properties:
for a 2D electron gas with quadratic dispersion, the Drude conductivity
depends on the effective mass $m$ of the electron, whereas in graphene
the same quantity depends on the mass $m_{g}$. Therefore it is permissible
to replace $m$ by $m_{g}$ in the hydrodynamic equation~\cite{key-7}.
A note is in order here: we have formulated the problem with the aid
of velocity fields which is a natural choice for massive particles.
Naturally, we could also have formulated the equation-of-motion in
terms of the momentum fields in which case we would have arrived at
the same final equation, but without the need to assign a Drude mass
to the electrons in the graphene.

%----------------------------------------------------------------------------------------------------------%

\end{document}